\documentclass[12pt]{iopart}

\usepackage{iopams}  

\usepackage[breaklinks=true,backref=none]{hyperref} 

\usepackage[utf8]{inputenc}

\usepackage{amsmath}
\usepackage{mathtools}
\DeclarePairedDelimiter\bra{\langle}{\rvert}
\DeclarePairedDelimiter\ket{\lvert}{\rangle}
\DeclarePairedDelimiterX\braket[2]{\langle}{\rangle}{#1 \delimsize\vert #2}

\usepackage[english]{babel}                      

\usepackage{ifthen}    

\usepackage{amsfonts,amssymb,amsthm,mathtools}    

\usepackage{wrapfig}
\usepackage{pictexwd}
\usepackage{xcolor}
\usepackage{mathtools}

\usepackage{hyperref}
\usepackage{float}
\usepackage{amsfonts}
\usepackage{mathrsfs}
\usepackage{psfrag}
\usepackage{indentfirst}
\usepackage{graphicx}
\usepackage{ragged2e}
\usepackage{caption}
\usepackage{subcaption} 
\usepackage{soul}

\usepackage{tikz}
\usetikzlibrary{arrows}
\usepackage{bm}
\usepackage{cite}

\usepackage[normalem]{ulem}

\begin{document}

\title{Time-dependent quantum geometric tensor and some applications}

\author{Bogar Díaz$^{1, 2}$, Diego Gonzalez $^{3, 4}$,
Marcos J. Hern\'andez$^{1}$, and
J. David Vergara$^{1}$\footnote{Corresponding author.}}
\address{$^1$ Departamento de F\'isica de Altas Energ\'ias, Instituto de Ciencias Nucleares, Universidad Nacional Aut\'onoma de M\'exico, Apartado Postal 70-543, Ciudad de M\'exico, 04510, M\'exico}
\address{$^2$ Group of Biometrics, Biosignals, Security and Smart Mobility (GB2S), E.T.S.I. Minas y Energía, Universidad Politécnica de Madrid, C. de Ríos Rosas, 21, 28003, Madrid, Spain}
\address{$^3$ Escuela Superior de Ingenier\'ia Mec\'anica y El\'ectrica, Instituto Polit\'ecnico Nacional, Unidad Profesional Adolfo L\'opez Mateos, Zacatenco, 07738 Gustavo A. Madero, Ciudad de M\'exico, M\'exico}
\address{$^4$ Departamento de F\'{i}sica, Cinvestav, Avenida Instituto Polit\'{e}cnico Nacional 2508, San Pedro Zacatenco, 07360 Gustavo A. Madero, Ciudad de M\'exico, M\'exico}
\ead{bogar.diaz@upm.es, dgonzalezv@ipn.mx, mhm@ciencias.unam.mx, vergara@nucleares.unam.mx}

\begin{abstract}
We define a time-dependent extension of the quantum geometric tensor to describe the geometry of the time-parameter space for a quantum state by considering small variations in both time and wave function parameters. Compared to the standard quantum geometric tensor, this tensor introduces new temporal components, enabling the analysis of systems with non-time-separable or explicitly time-dependent quantum states and encoding new information about these systems. In particular, the time-time component of this tensor is related to the energy dispersion of the system. We applied this framework to a harmonic/inverted oscillator, a time-dependent harmonic oscillator, and a chain of generalized harmonic/inverted oscillators. We show some results on the scalar curvature associated with the time-dependent quantum geometric tensor and the generalized Berry curvature behavior on the transition from harmonic oscillators to inverted ones. Furthermore, we analyze the entanglement for the chain through purity analysis, obtaining that the purity for any excited state is zero in the mentioned transitions.
\end{abstract}


\section{Introduction}\label{sec:level1}

The geometrical description of quantum phenomena has led to significant results in physics. Notably, Pancharatnam's work on polarized light interference \cite{Pancha} laid the foundation for understanding geometric phases. Independently, Berry \cite{Berry1984} and Simon \cite{Simon} showed that the Berry phase is a holonomy in a Hermitian line bundle, within the context of an adiabatic evolution. After that, these works were extended to consider non-adiabatic cases by Aharonov and Anandan \cite{AA1, AA2}, while Samuel and Bhandari \cite{Samuel, Samuel1} generalized them further to non-cyclic and non-unitary evolutions. Anandan and Aharonov \cite{AA3} approached the problem geometrically, showing that the energy uncertainty integral corresponds to the distance along a curve measured by the Fubini-Study metric. Subsequently, the work of Provost and Valle \cite{Provost} on the construction of the quantum geometric tensor (QGT) allowed them to give an utterly geometrical structure to the parameter space of a quantum system. Following the ideas of Provost and Valle, it has been shown that the geometry of the parameter space encodes much of the relevant quantum information of a system, such as the identification of phase transitions \cite{Zanardi2007Information, Carollo2020, Sarkar2012, Gonzalez2020}, along with non-equilibrium steady-state quantum phase transitions \cite{Tong, Marmo} and the excited state quantum phase transitions \cite{Leyvraz, Caprio}. 

Recent experiments have shown that both the real and imaginary parts of the QGT can be measured. For instance, complete Bloch-state tomography in an optical Raman lattice of ultracold atoms has been used to reconstruct the full QGT, both the quantum metric and Berry curvature, across the Brillouin zone \cite{Yi2023}. In solid-state spin systems, in Ref. \cite{Yu2020}  using a qubit formed by a Nitrogen Vacancy center in diamond, they performed an experimental measurement of metric and curvature components of the QGT. More generally, periodic driving schemes—where one modulates a Hamiltonian parameter and measures the resulting excitation rate—yield each metric element via the linear response of transition probabilities; this approach has been demonstrated, for instance, in a superconducting qubit array \cite{Ozawa2018}.

In this work, we extend and unify the previous theoretical developments by considering a non-adiabatic evolution and a manifold constructed out of the parameters and the system's time dependence. In this way, we carry out a covariant construction of the QGT, a time-dependent quantum geometric tensor (tQGT), in the sense that we consider the parameters and time on the same foot. This is analogous to a relativistic space-time description, but with a Euclidean signature. A notable feature of our formalism is that in the case of stationary states, the temporal components of the tQGT are zero, and our development is directly reduced to the QGT of Provost and Valle \cite{Provost}. We find that the Berry curvature is generalized, called tBerry curvature, and acquires time components containing non-trivial information about the system and can be thought of as the components of an ``electric field".
This allows to obtain information from the system that can not be obtained in any other way. This is illustrated through the harmonic oscillator, where the tQGT enables us to identify the ground state of the system, corresponding to the place where the time-parameter components of the tBerry curvature change sign or the time-parameter components of the generalized metric become zero. The tQGT also allows us to identify the transition from a harmonic oscillator to an inverted harmonic oscillator. In addition, whether the energy of the inverted oscillator is greater than zero or less than zero. 

This paper is organized as follows. In Sec. \ref{sec2}, we introduce the tQGT ($Q_{IJ}$) and divide it into real and imaginary parts. The real part of the new tensor defines a time-dependent quantum metric that generalizes the quantum metric tensor, while its imaginary part generalizes the Berry curvature. We show that the tQGT is gauge invariant under time-dependent gauge transformations and that the time-time component ($Q_{00}$) corresponds to the energy dispersion. Furthermore, we prove that the tQGT reduces to the standard QGT for stationary states. Section \ref{sec3} introduces our first example, the harmonic and inverted oscillators. Both systems are studied using Gaussian solutions that are non-separable in time, allowing us to give a unified description. We analyze the general case of these solutions and build the tQGT, showing what information can be obtained. We also study in detail the transition zone between harmonic and inverted, displaying that the tQGT diverges at the transition point. Section~\ref{secHOTDF} contains an example of a system with a time-dependent Hamiltonian. To be precise, we study the tQGT for a time-dependent harmonic oscillator.
In Sec.~\ref{sec:coupledsystems} we extend the results of Sec. \ref{sec3} to a system of coupled particles. Furthermore, considering several degrees of freedom and the fact that our states are Gaussian, we also compute and analyze the purity of the quantum system of coupled particles. We pay special attention to the transition zone, where we show that the purity of the reduced system is zero for all even excited eigenstates. Finally, we give our conclusions in Sec. \ref{sec:conclu}.

\section{Time-dependent quantum geometric tensor}\label{sec2}

In this section, we introduce a tQGT. We begin by considering a system, defined by a Hamiltonian operator $\hat{H}$, in a quantum state at time $t$, represented by a normalized state $\ket{\Psi(t;\lambda)}$ with a set of parameters $\lambda=\{\lambda^i\}$ ($i,j,\dots\!=1,\dots,\mathcal{N}$). We assume that time and parameters are independent quantities, and then consider a full set of time and parameters denoted by $\Lambda=\{\Lambda^I\}=\{\Lambda^0,\Lambda^i\}$ ($I,J,\dots\!=0,1,\dots,\mathcal{N}$) with $\Lambda^0=t$ and $\Lambda^i=\lambda^i$. Using this notation, we can write the state of the system as $\ket{\Psi(\Lambda)}:=\ket{\Psi(t;\lambda)}$. Notice that, in general, the Hamiltonian can be time-dependent and can depend on the parameters $\lambda$, but not necessarily all of them.

Let us now introduce an infinitesimal displacement $\Lambda \rightarrow \Lambda'=\Lambda+\delta \Lambda$, and the corresponding ket state at $\Lambda'$, which we denote as $\ket{\Psi(\Lambda+\delta \Lambda)}$. The distance between the states $\ket{\Psi(\Lambda)}$ and $\ket{\Psi(\Lambda+\delta \Lambda)}$ is defined by 
\begin{equation}\label{QGTdistance}
dl^2\equiv 1 - | \braket{\Psi(\Lambda)}{\Psi(\Lambda+\delta \Lambda)} |^2 \,,
\end{equation}
where $| \braket{\Psi(\Lambda)}{\Psi(\Lambda+\delta \Lambda)} |$ is recognized as the fidelity. 

Expanding the state $\ket{\Psi(\Lambda+\delta \Lambda)}$ into a second-order Taylor series, we have
\begin{align}\label{taylor}
    \ket{\Psi(\Lambda+\delta \Lambda)} \simeq   &\ket{\Psi(\Lambda)}+ \sum_{I=0}^{\mathcal{N}} \, \ket{\partial_I\Psi(\Lambda)} \, \delta \Lambda^I +\frac12 \sum_{I,J=0}^{\mathcal{N}} \,\ket{\partial_I\partial_J\Psi(\Lambda)}\, \delta \Lambda^I \delta \Lambda^J  \,,
\end{align}
where $\partial_I\equiv \frac{\partial}{\partial \Lambda^I}$. Using \eqref{taylor}, it is straightforward to see that the distance $\eqref{QGTdistance}$ can be expressed as
\begin{equation}
 dl^2\simeq \sum_{I,J=0}^{\mathcal{N}} \, g_{IJ}(\Lambda) \, \delta \Lambda^I \, \delta \Lambda^J\,, 
\end{equation}
where we designate the time-dependent quantum metric tensor (tQMT) as
\begin{equation}\label{metric-tensor-time}
    g_{IJ}(\Lambda) :={\rm Re}(Q_{IJ}(\Lambda))\,,
\end{equation}
which is the real part of the Hermitian tensor
\begin{align}\label{QGT-time}
Q_{IJ}(\Lambda) :=& \braket{\partial_I\Psi(\Lambda)}{\partial_J\Psi(\Lambda)} -\braket{\partial_I\Psi(\Lambda)}{\Psi(\Lambda)}\braket{\Psi(\Lambda)}{\partial_J\Psi(\Lambda)}\,.
\end{align}
We define $Q_{IJ}(\Lambda)$ as the tQGT and it can be regarded as a generalization of the standard QGT, as it includes new components associated with the time index ($I=0$), specifically $Q_{00}$, $Q_{0i}$ and $Q_{i0}$. Furthermore, an important feature of the tensor $Q_{IJ}(\Lambda)$ is that it provides new and nontrivial information when $\ket{\Psi(\Lambda)}$ is not separable in time. On the other hand, the tQMT, $g_{IJ}(\Lambda)$, plays a fundamental role in describing the geometrical aspects of the parameter space, since it allows the measurement of the distance between the nearby states $\ket{\Psi(\Lambda)}$ and $\ket{\Psi(\Lambda+\delta \Lambda)}$ in the time parameter manifold $\mathcal{M}$. 

The tensor $Q_{IJ}(\Lambda)$ is in general complex, and its imaginary part can be associated with a 2-form time-dependent curvature $F=\frac12 \sum_{I, J=0}^{\mathcal{N}} \, F_{IJ} \delta \Lambda^I \wedge  \delta \Lambda^J$ with components
\begin{equation}\label{Berry-tensor-time}
    F_{IJ}(\Lambda)=-2{\rm Im}(Q_{IJ}(\Lambda))\,.
\end{equation}
Further, these components can be written as $F_{IJ}(\Lambda)=\partial_I A_J(\Lambda) - \partial_J A_I(\Lambda)$, where $A_I(\Lambda)$ are the time-dependent connection coefficients
\begin{equation}\label{connection-time}
    A_I(\Lambda)= -{\rm Im}\left( \braket{\Psi(\Lambda)}{\partial_I\Psi(\Lambda)} \right)\,.
\end{equation}
It is clear that \eqref{connection-time} can be thought of as a generalization of the Berry connection \cite{Berry1984}. For this reason, we call $F$ the tBerry curvature.

On the other hand, it is not hard to show that under the gauge transformation
\begin{equation}
\ket{\Psi(\Lambda)} \rightarrow \ket{\Psi'(\Lambda)}=\exp{(\rm i \alpha(\Lambda))} \,  \ket{\Psi(\Lambda)}\,, \label{gauge} 
\end{equation}
where $\alpha(\Lambda)$ is an arbitrary real function of time and parameters, the tQGT $Q_{IJ}(\Lambda)$ is gauge-invariant, i.e.,
\begin{equation}
    Q_{IJ}(\Lambda) \rightarrow Q'_{IJ}(\Lambda) = Q_{IJ}(\Lambda)\,,
\end{equation}
while the time-dependent connection $A_I(\Lambda)$ transforms as
\begin{equation}\label{connection-time-gauge}
  A_I(\Lambda) \rightarrow A'_I(\Lambda) = A_I(\Lambda) - \partial_I \alpha(\Lambda)\,,
\end{equation}
i.e., $A_I(\Lambda)$ is gauge-dependent.

It worth noting that assuming the time evolution of $\ket{\Psi(\Lambda)}$ is determined by the Schr\"{o}dinger equation, $\hat{H}\ket{\Psi(\Lambda)}={\rm i} \hbar \ket{\partial_0 \Psi(\Lambda)}$, the temporal component of \eqref{QGT-time} becomes
\begin{align}
    Q_{00}(\Lambda) &= \frac{1}{\hbar^2} \left( \bra{\Psi(\Lambda)} \hat{H}^2 \ket{\Psi(\Lambda)} - \bra{\Psi(\Lambda)} \hat{H} \ket{\Psi(\Lambda)}^2 \right) \nonumber\\
    &= \frac{(\Delta E)^2}{\hbar^2}\,, \label{Q00}
\end{align}
where $(\Delta E)^2$ is the energy dispersion. Equation~\eqref{Q00} reproduces the result obtained by Anandan and Aharonov in~\cite{Anandan_1990}, where only the infinitesimal time displacement $t \rightarrow t'=t+\delta t$  is taken into account. Furthermore, in this case the time-parameter components of \eqref{QGT-time} reduce to
\begin{align}
    Q_{0i}(\Lambda) &= \frac{{\rm i}}{\hbar} \left( \bra{\Psi(\Lambda)} \hat{H} \ket{\partial_i \Psi(\Lambda)} - \bra{\Psi(\Lambda)} \hat{H} \ket{\Psi(\Lambda)} \braket{\Psi(\Lambda)}{\partial_i\Psi(\Lambda)} \right)\nonumber\\
    &=\frac{{\rm i}}{\hbar}\left[ \bra{\Psi(\Lambda)} (\hat{H} - \langle \hat{H}\rangle ) \ket{\partial_i \Psi(\Lambda)} \right],\label{Q0i}
\end{align}
where $\langle \hat{H}\rangle = \bra{\Psi(\Lambda)} \hat{H} \ket{\Psi(\Lambda)}$. From this, it follows that the time-parameter components of the tQMT and the tBerry curvature are
\begin{align}
    g_{0i}(\Lambda) = -\frac{1}{\hbar} {\rm Im}\left[ \bra{\Psi(\Lambda)} (\hat{H} - \langle \hat{H}\rangle ) \ket{\partial_i \Psi(\Lambda)} \right], \label{tQMT0i}
\end{align}
and  
\begin{align}
    F_{0i}(\Lambda) &= -\frac{2}{\hbar}  {\rm Re}(\bra{\Psi(\Lambda)} \hat{H} \ket{\partial_i \Psi(\Lambda)}) \nonumber \\
    &= \frac{1}{\hbar} \left( \bra{\Psi(\Lambda)} \partial_i \hat{H} \ket{\Psi(\Lambda)} - \partial_i \langle \hat{H}\rangle \right), \label{tBerry0i}
\end{align}
respectively. Thus, we observe that these components of the tBerry curvature measure the difference between the expected value of the Hamiltonian change concerning the parameters and the change of the expected value.
To introduce a more physical interpretation of these components of the tQGT, we consider that $\hat P_I$ are Hermitian operators that generate the infinitesimal displacement  
\begin{equation}
    |\Psi(\Lambda')\rangle =\exp\left(-\frac{i}{\hbar} \delta \Lambda^I \hat P_I \right)  |\Psi(\Lambda)\rangle\,, 
\end{equation}
which, to first-order, we write as 
\begin{equation}
    i\hbar |\partial_I \Psi(\Lambda)\rangle = \hat P_I|\Psi(\Lambda) \rangle.
\end{equation}
In this form, we can rewrite the components of the tQMT \eqref{Q00} and \eqref{tQMT0i} in compact form as
\begin{equation}
      g_{0I}(\Lambda) = \frac{1}{\hbar^2} \left[\frac{1}{2} \langle\{\hat{H} ,\hat P_I\}_+ \rangle - \langle \hat{H}\rangle \langle \hat P_I \rangle \right],
\end{equation}
where $\{,\}_+$ denotes the anticommutator. This expression is quite similar to the one obtained in the path integral representation of the QMT \cite{Alvarez2017} and it reduces to the expression \eqref{Q00} in the case of $\hat{P}_I=\hat{P}_0=\hat{H}$. Furthermore, it indicates the proportion by which the expected value of the operators' anticommutator differs from the product of the expected values of the operators, which corresponds to a covariance matrix of these operators. For the tBerry curvature we obtain from \eqref{tBerry0i} 
\begin{equation}
  F_{0i} = \frac{2}{\hbar^2} \langle \Psi(\Lambda) | [\hat H,\hat P_i] | \Psi(\Lambda) \rangle.
\end{equation}
In this way, the Berry curvature tells us the proportion in which the Hamiltonian commutes with the generator of translations of the parameters.
In addition, the temporal component of \eqref{connection-time} takes the form
\begin{equation}\label{A0}
    A_0(\Lambda)=\frac{1}{\hbar} \langle \hat{H}\rangle,
\end{equation}
meaning that $A_0$ is proportional to the energy expectation value.

Notice that for a stationary state
\begin{equation}\label{eigenkets}
    \ket{\Psi(\Lambda)}=\exp{\left(-\frac{{\rm i} E_n(\lambda)}{\hbar} t \right)} \, \ket{n(\lambda)} \,,
\end{equation}
where $\ket{n(\lambda)}$ satisfy the eigenvalues equation $\hat{H}\ket{n(\lambda)}=E_n(\lambda) \ket{n(\lambda)}$ with nondegenerate eigenvalue $E_n(\lambda)$, equation~\eqref{Q00} reduces to  $Q_{00}(\Lambda)=0$, while \eqref{A0} leads to $A_0(\lambda)=E_n(\lambda)/\hbar$. Therefore, the component $Q_{00}(\Lambda)$ can be interpreted as a quantitative measure of the system’s non-stationarity, while $A_0(\Lambda)$ is the gauge potential along the temporal coordinate, which gives rise to the dynamical phase acquired by the state. Moreover, for the stationary state \eqref{eigenkets}, the only non-vanishing components of \eqref{QGT-time} are the parameter components 
\begin{equation}\label{Qstationary}
    Q_{ij}(\lambda)=\braket{\partial_i n(\lambda)}{\partial_j n(\lambda)} -\braket{\partial_i n(\lambda)}{n(\lambda)}\braket{n(\lambda)}{\partial_j n(\lambda)}\,,
\end{equation}
which do not depend on time and can be recognized as the components of the standard QGT~\cite{Provost}.

In what follows, we explore the consequences of \eqref{QGT-time} by considering general quantum states that are not stationary.

\section{Harmonic oscillator and inverted harmonic oscillator}\label{sec3}

Along this paper, we work with solutions of the time-dependent Schr\"{o}dinger equation that are non-time-separable. In the case of one dimension, we consider the family of Gaussian solutions of the form
\begin{align}
\psi(q,\Lambda) =C(\Lambda)e^{-(U(\Lambda)+iV(\Lambda))q^2/2 }\,,
	\label{eq:funcionondagen}
\end{align}
where  $U(\Lambda)>0$, $V(\Lambda)\in \mathbb{R}$, and $C(\Lambda)\in \mathbb{C}$. We remember that $\Lambda= \{t, \lambda^i\}$. Hereafter, we use $\hbar=1$. The normalization condition on \eqref{eq:funcionondagen} implies
\begin{align}         
|C(\Lambda)|^2=\sqrt{\frac{U(\Lambda)}{\pi}}\,.
\end{align}

Using the approach introduced in Section \ref{sec2} for the calculation of the tQGT and taking the real part, we find that the tQMT components  \eqref{metric-tensor-time} are given by
\begin{align}
g_{IJ}=\frac{\partial_{I}U\partial_{J}U+\partial_{I}V\partial_{J}V}{8U^2}\,.
	\label{gijgeneral}
\end{align}
On the other hand, taking the imaginary part, we find that the components of the tBerry curvature \eqref{Berry-tensor-time} correspond to
\begin{align}
	F_{IJ}=\frac{-\partial_{I}U\partial_{J}V+\partial_{I}V\partial_{J}U}{4U^2}\,.
	\label{fijgeneral}
\end{align}

For this model, the determinant of the metric is zero if $\mathcal{N}>2$, which implies that only two elements $\Lambda^I$ can be independent. Therefore, in the following, we will only consider the metrics obtained by varying any two selected elements, $\Lambda^{\scriptscriptstyle I_1}$ and $\Lambda^{ \scriptscriptstyle I_2}$, while keeping the rest fixed. We denote this metric by $\tilde{\boldsymbol{g}}[\Lambda^{\scriptscriptstyle I_1},\Lambda^{\scriptscriptstyle I_2}]$, explicitly, it is given by
\begin{align}
\tilde{\boldsymbol{g}}  [\Lambda^{\scriptscriptstyle I_1},\Lambda^{\scriptscriptstyle I_2}]=\left(
	\begin{array}{cc}
		g_{_{I_1 I_1}} & g_{_{I_1 I_2}}  \\
		g_{_{I_1 I_2}} & g_{_{I_2 I_2}} 
	\end{array}
	\right)\,.
\end{align}

All these metrics, for any two selected parameters, satisfy the Palumbo and Goldman relation \cite{Palumbo}, which was first observed in the context of the dimensional reduction in \cite{Freund, Nepo}. Specifically, the relation is $ \det \tilde{\boldsymbol{g}}[\Lambda^{{\scriptscriptstyle I_1}},\Lambda^{{\scriptscriptstyle I_2}}]= \left(F_{\scriptscriptstyle I_1 I_2}\right)^2$, with $I_1$ and $I_2$ fixed. On the other hand, the scalar curvature for any of these metrics is denoted by $\mathcal{R}[\Lambda^{I_1},\Lambda^{I_2}]$ and satisfies $\mathcal{R}[\Lambda^{I_1},\Lambda^{I_2}]=-16$ if $\det \Tilde{\boldsymbol{g}}[\Lambda^{{\scriptscriptstyle I_1}},\Lambda^{{\scriptscriptstyle I_2}}]\neq 0$, otherwise it cannot be calculated. This means that the 2-dimensional time-parameter space of
these Gaussian states possess a hyperbolic geometry.
These properties of the metric for systems with wave function \eqref{eq:funcionondagen} are summarized in the following set of equations
\begin{subequations}
\begin{align} \displaystyle
    \det \boldsymbol{g}&= 0 \,,  \label{eq:determinante0} \, &\mathrm{if} \, \,  \mathcal{N}>2\,,\\  
   \det\tilde{ \boldsymbol{g}}[ \Lambda^{\scriptscriptstyle I_1},\Lambda^{ \scriptscriptstyle I_2}]&= \frac{1}{4}\left(F_{\scriptscriptstyle I_1 I_2} \right)^2 \,, \,& \mathrm{if} \, \,  \mathcal{N}=2\,,
    \label{eq:palumbo}\\
    \mathcal{R}[\Lambda^{I_1},\Lambda^{I_2}]&=-16  \,, & \mathrm{if} \, \, \mathcal{N}=2\label{eq:curvatura1d} \, \, \mathrm{and} \, \, \det \tilde{\boldsymbol{g}}[\Lambda^{{\scriptscriptstyle I_1}},\Lambda^{{\scriptscriptstyle I_2}}]\neq 0 \,.
\end{align}
\label{eq:propiedadesgenerales}
\end{subequations}

A system that admits this kind of solution is
\begin{align}
\hat{H}(\hat{q},\hat{p}, \Lambda)&=\frac{1}{2} \left( W \hat{p}^2 + X\hat{q}^2\right)\,, \label{firstHamil}
\end{align}
where $W>0$ and $X\in \mathbb{R}$. Notice that this system presents a bifurcation at $X=0$. From the classical point of view, for $X>0$, the system corresponds to the simple harmonic oscillator, where we have an elliptical stable fixed point at $(p, q)=(0,0)$. On the other hand, for $X<0$, the system represents an inverted harmonic oscillator, with a hyperbolic unstable fixed point at $(p, q)=(0,0)$. We refer to $X>0$ and $X<0$ as regions 1 and 2, respectively. The simple and the inverted harmonic oscillators conserve energy, making them integrable. From a quantum point of view, the case $X>0$ has normalizable and time-separable solutions, given by the well-known Hermite functions. However, for $X<0$, such solutions do not exist \cite{Barton}. To obtain normalizable solutions, it is possible to follow two strategies: confine the system as in \cite{Yuce} or consider solutions that are non-time-separable as in \cite{Ali}. For our purposes, the latter is more convenient and the one we will follow in this work. Notably, this type of solution also exists for the case of $X>0$, as seen in \cite{Griffiths}. 

In what follows, we consider the cases $X>0$ and $X<0$ separately, and then we compare the information obtained from the two systems and analyze the transition of the two regions. The idea is to start in both regions with an initial state at $t=0$,
\begin{align}
    \psi(q, 0; \lambda)=\left( \frac{B \sqrt{ |XW| }}{W \pi} \right)^{1/4} \exp \left(- \frac{ \sqrt{ |XW| } B q^2}{2 W}\right)\,,
    \label{inicial}
\end{align}
where $\{\lambda^i\}=\{X,W,B\}$, with $i=1,2,3$ and $B>0$. Later, we evolve this initial state over time with the system's kernel. We then analyze the resulting wave functions using the tQGT to see what information we observe from the system.

\subsection{Region 1: harmonic oscillator}

In this case, the kernel of the system is given by
\begin{align}\label{eq:kernel}
K(q',t;q,0)= \sqrt{\frac{\omega}{2\pi i W t}}\exp{\left(-\frac{\omega((q^2+q'^2)\cos \left( \omega t \right) -2qq')}{2iW\sin \left( \omega t \right)}\right)}\,,
\end{align}
where $\omega=\sqrt{ XW }$. Using \eqref{eq:kernel}, the wave packet \eqref{inicial} takes the form
\begin{align}
\psi_{HO}(q,t;\lambda)=\left( \frac{B \omega}{W \pi} \right)^{1/4} \frac{1}{\sqrt{\cos \left( \omega t \right)+i
   B \sin \left( \omega t \right)}}  \exp \left(- \frac{q^2}{2}\frac{ \omega\left(1-i B \cot\left( \omega t \right)\right)}{W \left(B-i \cot \left( \omega t \right)\right)}\right)\,.
   \label{eq:funciononda}
\end{align}
 This is a particular case of the wave function \eqref{eq:funcionondagen}, where the functions $U(\Lambda)$, $V(\Lambda)$ and $C(\Lambda)$ are given by
\begin{subequations}
\begin{align}
    U_{HO}(\Lambda)&=\frac{B X \csc^2 \left( \omega t \right)}{\omega  \left(B^2+\cot ^2 \left( \omega t \right) \right)}\,,\\
    V_{HO}(\Lambda)&= \frac{\left(B^2-1\right) X \cot \left( \omega t \right)}{\omega  \left(B^2+\cot^2 \left( \omega t \right)\right)}\,,\\
    C_{HO}(\Lambda)&=\left( \frac{B \omega}{W \pi} \right)^{1/4} \frac{1}{\sqrt{\cos \left( \omega t \right)+i
   B \sin \left( \omega t \right)}}\,.
\end{align}
\label{UVCHO}
\end{subequations}

Note that, for $B=1$, \eqref{eq:funciononda} reduces to the usual ground state wave function of the harmonic oscillator. This suggests that the parameter $B$ must be associated with the system's average energy. To show this, we calculate the expected value of the Hamiltonian \eqref{firstHamil} on the state \eqref{eq:funciononda}. The result is
\begin{align}\label{eq:energy}
\langle \hat{H}\rangle =\frac{(1+B^2)\omega}{4B}\,.
\end{align}
From \eqref{eq:energy}, it is clear that the minimum energy corresponds to  $B=1$, which is precisely the energy of the ground state. 

On the other hand, since $B$ is a continuous parameter, for $B\neq 1$, the state \eqref{eq:funciononda} corresponds to a linear combination of energy basis states. To show this, we begin by writing \eqref{eq:funcionondagen} in terms of this basis, which results in 
\begin{align}\label{27}
    \Psi(q,t;\lambda)
    =&\sum_{n=0}^{\infty} \exp{(-iE_nt)}u_n(q) \langle n | \Psi(0;\lambda)  \rangle \nonumber  \\ \vspace*{-2cm} \nonumber \\
    =&\sum_{n=0}^{\infty} \frac{1}{\sqrt{(2n)!\mu\gamma}} \exp{(-iE_{2n}t)}u_{2n}(q)  C(0;\lambda)2^{n-1/2} \Gamma(n+1/2)\left(1- \frac{2}{\mu} \right)^n\,,
\end{align}
where $E_n$ are the energy eigenvalues of the system, $u_n(q)=\langle q|n\rangle $ and

\begin{align}    
\mu&=1+\frac{U(0;\lambda)+iV(0;\lambda)}{\gamma^2}\,\label{un}
\end{align}
with $\gamma=\sqrt{\omega/W}$. Substituting \eqref{UVCHO} in \eqref{27} and taking 
$u_n(q)=a_n\exp{\left(-\frac{\gamma^2 q^2}{2}\right)}H_n\left( \gamma q \right)\,$  we obtain for the Harmonic oscillator 
\begin{align}
\psi_{HO}&(q,t;\lambda) = \frac{B^{1/4}}{\sqrt{1+B}}\sum_{n=0}^{\infty}  \frac{e^{-iE_{2n}t}\sqrt{(2n)!}}{2^{n+1/2}n!}\left(\frac{1-B}{B+1}
     \label{fourierpart} \right)^n u_{2n}(q)  \,, 
\end{align}
which is the expansion of \eqref{eq:funciononda} in terms of energy basis states. From this expression, we can see that $\psi_{HO}(q,t;\lambda)$ is the ground state for $B=1$, while for all other values of $B$, it is a linear combination of the even energy eigenstates. 

We can also notice that for a given value of the average energy $\langle \hat{H} \rangle= \bar E$, Eq.~\eqref{eq:energy} has two solutions for the parameter $B$, which are given by
\begin{equation}\label{bee1}
    B^{\pm} = \frac{2\bar E \pm \sqrt{4 \bar E^2-\omega^2 }}{\omega}\,.
\end{equation}
Notice that $B^{+}>1$, $B^{-}<1$, and $B^+ B^- =1$. Also, for the ground state, $B^{+}=B^{-}=1$.

We now take $\Lambda=\{\Lambda^I\}= (t, X, W, B)$ with $I=0,1,2,3$ as the varying parameters in the wave function \eqref{eq:funciononda} and proceed to compute the tQGT.

Using  \eqref{UVCHO} together with  \eqref{gijgeneral}, we obtain the components of the tQMT
\begin{subequations} \allowdisplaybreaks
\begin{align}
g_{00}&=\frac{\left(B^2-1\right)^2\omega^2}{8 B^2} \label{g00}\,,\\
g_{01}&=-\frac{\left(B^2-1\right) W \left(\left(B^2+1\right) \sin \left(2  \omega t\right)-2 \left(B^2-1\right)  \omega t \right)}{32 B^2 \omega} \,,\\
g_{02}&=\frac{\left(B^2-1\right) X \left(2 \left(B^2-1\right)  \omega t+\left(B^2+1\right) \sin \left(2  \omega t \right)\right)}{32 B^2 \omega}\,, \\
g_{03}&=0 \,, \\
g_{11} &=\frac{-8 \left(B^4-1\right)  \omega t  \sin \left(2  \omega t \right)+B^4+8 \left(B^2-1\right)^2 \left( \omega t \right)^2 -\left(B^2-1\right)^2 \cos \left(4  \omega t \right)+6 B^2+1}{256 B^2 X^2}\,,\\
g_{12} &=-\frac{B^4-8 \left(B^2-1\right)^2 \left( \omega t \right)^2 -\left(B^2-1\right)^2 \cos \left(4 \omega t \right)+6 B^2+1}{256 B^2 \omega^2 }\,,\\
\label{g22}
g_{22} &=\frac{8 \left(B^4-1\right)  \omega t \sin \left(2  \omega t\right)+B^4+8 \left(B^2-1\right)^2 \left( \omega t \right)^2-\left(B^2-1\right)^2 \cos \left(4  \omega t \right)+6 B^2+1}{256 B^2 W^2}\,,\\
g_{13} &=\frac{\cos \left(2  \omega t \right)}{16 B X}\,,\\
g_{23} &=-\frac{\cos \left( 2 \omega t \right)}{16 B W}\,,\\
g_{33} &=\frac{1}{8 B^2}\,.
\end{align}
\end{subequations}

Some remarks about these metric components. First, notice that for both values of the parameter $B$ given by \eqref{bee1}, the metric component \eqref{g00} reduces to the same energy dispersion $g_{00}= \frac{2 \bar E^2}{\omega}-\frac{\omega}{2}$. Also, all metric components diverge when $B \to 0$, which corresponds to infinite average energy \eqref{eq:energy}. Furthermore, for $B=1$ all the components of the QMT are bounded, and the metric $\tilde{\boldsymbol{g}}[X,W]$ reduces to
\begin{align}\label{QMT_B1}
\tilde{\boldsymbol{g}}[X,W]_{B=1}=\left(
\begin{array}{cc}
\frac{1}{32X^2} & -\frac{1}{32XW}  \\
 -\frac{1}{32XW} & \frac{1}{32W^2}
\end{array}
\right)\,,
\end{align}
which corresponds to the quantum metric tensor of the ground state of the harmonic oscillator \cite{Juarez}. Lastly, for an arbitrary $B$ ($\neq1$), the components of the QMT with time dependence (excepting $g_{13}$ and $g_{23}$) are unbounded as $ \omega t \to \infty$.

Now, we analyze the tQMT components' behavior, giving particular attention to large $\omega t$.  We can distinguish two different behaviors.

\begin{figure}[H]
  \centering
  \begin{subfigure}{0.45\linewidth}
    \includegraphics[width=\textwidth]{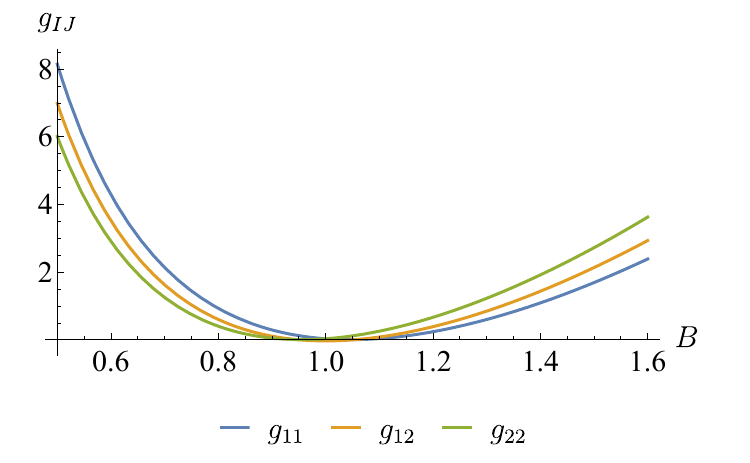}
    \caption{}
    \label{fig:gvariandobIa}
  \end{subfigure}
  \begin{subfigure}{0.45\linewidth}
    \includegraphics[width=\textwidth]{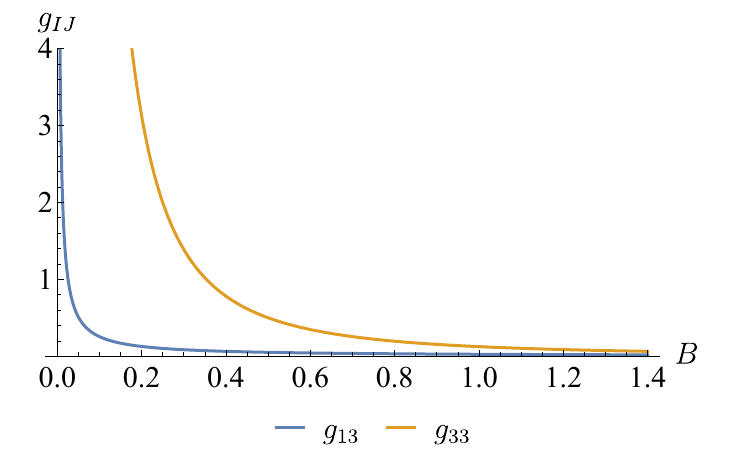}
    \caption{}
    \label{fig:gvariandobIb}
  \end{subfigure}
  \captionsetup{font=small} 
  \caption{\justifying Some components of the tQMT, as a function of the parameter $B$. Except for $g_{03}$, the tQMT components are singular at $B=0$. The remaining parameters were fixed as follows: $X=W=1$ and $t=10$. (a) These components have a minimum at around $B=1$. (b) These components show a monotone behavior.}
  \label{fig:gvariandobI}
\end{figure}

\begin{enumerate}
 \item The components $g_{00}$ and $g_{12}$ attain a minimum at $B=1$; while $g_{11}$ and $g_{22}$ have a minimum at $B=B_{H+}$ and $B=B_{H-}$, respectively, where 
\begin{align} \label{BHpm}
    B_{H\pm}&=\sqrt{\frac{2\omega t \pm \sin\left(2 \omega t\right)} {2 \omega t \mp\sin\left(2 \omega t\right)}}\,;
\end{align}
and finally, $g_{01}$ and $g_{02}$ have a minimum at $B=\sqrt{B_{H+}}$ and $B=\sqrt{B_{H-}}$, respectively. Notice that $B_{H\pm}\to 1$ when $\omega t\to \infty$. In fact, equation \eqref{BHpm} can be written as $B_{H\pm}=1\pm\Delta B_H$.  Hence, `the vicinity of $B=1$' in this context means the interval $[ 1-|\Delta B_H|,1+|\Delta B_H| ]$, with $\Delta B_H\ll 1$. Neglecting second-order terms on $\Delta B_H$, we find $\Delta B_H \approx \tfrac{\sin 2\omega t}{2\omega t}$. From this, it is clear that as $\omega t$ increases, $B_{H\pm}$ approaches to 1. Then, the components \eqref{g00}-\eqref{g22} have a minimum in the vicinity of $B=1$, corresponding to the ground state. This shows that our tQMT distinguishes the ground state of the system.  This behavior can be appreciated in Fig. \ref{fig:gvariandobIa}, where we take $X=W=1$ and $t=10$. 
 \item  The components $g_{13}$, $g_{23}$, and $g_{33}$ have a monotonic decreasing behavior, as we can see in Fig. \ref{fig:gvariandobIb}.
\end{enumerate}

\begin{figure}[ht!]
    \centering
    \includegraphics[width=0.55\linewidth]{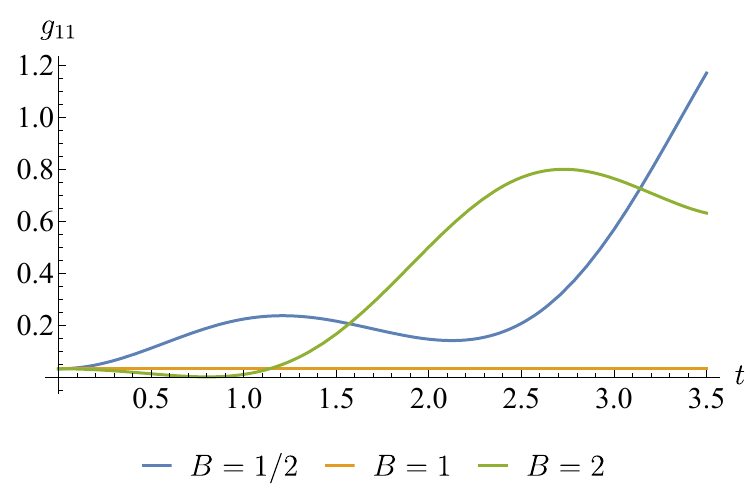}
  \captionsetup{font=small} 
    \caption{\justifying Component $g_{11}$ of the tQMT in the region 1 as a function of time. The parameters were fixed as $X=W=1$.}
    \label{g11osc}
\end{figure}

On the other hand, Fig. \ref{g11osc} shows the time evolution of the component $g_{11}$ for three values of the parameter $B$. For $B=1$, $g_{11}$ is time-independent. This is because the metric reduces to that of the ground state for this value (see \eqref{QMT_B1}). For $B=1/2$ and $B=2$, $g_{11}$ oscillates and increases, indicating that for variations of $X$ the distance between the states increases.  

\begin{figure}[H]
    \centering
    \includegraphics[width=0.55\linewidth]{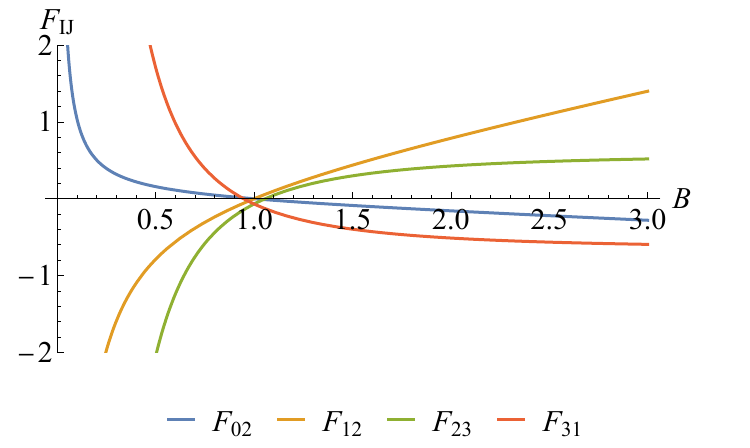}
    \captionsetup{font=small} 
    \caption{ Non-zero components of the tBerry curvature in the region 1 by fixing $X=W=1$. As a function of the parameter $B$, taking $t=5$. They cross zero around $B=1$. }
    \label{berry01}
\end{figure}

It is remarkable that, in contrast to time-separable states, the wave function \eqref{eq:funciononda} leads to a tBerry curvature different from zero. Using \eqref{fijgeneral} and \eqref{UVCHO}, the components of this curvature are
\begin{subequations}
    \begin{align}
    F_{01}&=\frac{\left(B^2-1\right) W \cos \left(2  \omega t\right)}{8 B \omega}\, ,\\
    F_{02} &= -\frac{\left(B^2-1\right) X \cos \left(2  \omega t \right)}{8 B \omega }\,,\\
    F_{03}&= \frac{\left(B^2-1\right) \omega}{4 B^2} \,,\\
     F_{12} &=-\frac{\left(B^2-1\right) t \cos \left(2  \omega t \right)}{8 B \omega }\,,\\ 
    F_{23} &=\frac{2 \left(B^2-1\right)  \omega t +\left(B^2+1\right) \sin \left(2  \omega t \right)}{16 B^2 W}\,,\\
    F_{31} &=\frac{\left(B^2+1\right) \sin \left(2  \omega t \right)-2 \left(B^2-1\right)  \omega t}{16 B^2 X}\,.
    \end{align}
\end{subequations}
In Fig. \ref{berry01}, we see that all the components of the curvature change sign in the vicinity of $B=1$, which, according to \eqref{fourierpart}, is the only value of $B$ that corresponds to a stationary state. More precisely, $F_{0i}$ and $F_{12}$, cross zero at $B=1$, because in that case we have a stationary state, while $F_{13}$ and $F_{23}$ cross zero at $B=B_{H+}$ and $B=B_{H-}$, respectively. 

\begin{figure}[H]
    \centering
    \includegraphics[width=0.55\linewidth]{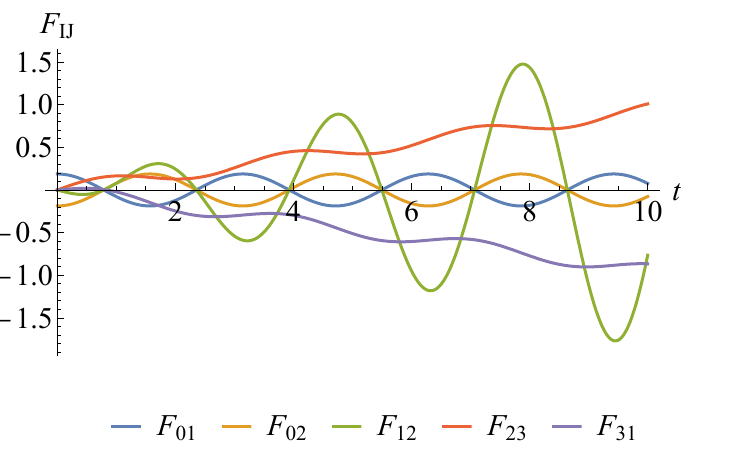}
    \captionsetup{font=small} 
    \caption{\justifying Oscillatory components of the tBerry curvature for the harmonic oscillator as a function of time, taking $B=2$ and $X=W=1$. }
    \label{berryo2}
\end{figure}

In Fig. \ref{berryo2} we plot the tBerry curvature components that oscillate in time, as we show in the next Subsection, this is related to the behavior of the probability distribution.

\subsubsection{Probability density}

We now analyze the behavior of the distribution probability $\rho_{HO}(q,t;B)=\psi_{HO} \psi^*_{HO}$, considering only the dependence on $q$, $t$, and $B$.  We begin by noting that, as a function of time, $\rho_{HO}(q,t;B)$ has critical points $(\partial_t\rho_{HO}=0)$ at $t^*=\tfrac{n\pi}{2\omega}$ with $n \in \mathbb{Z}$. In particular, the second time derivative of $\rho_{HO}$ evaluated at $q=0$ and  $t=t^*$ gives
\begin{subequations}\label{secondderivative}
\begin{align}
\partial^2_t\rho_{HO}(0,t^*_{\text{even}};B)&=-(B^2-1)\sqrt{\frac{B\omega^{5}}{W\pi}}\,,\\
     \partial^2_t\rho_{HO}(0,t^*_{\text{odd}};B)&=\frac{B^2-1}{B^3}\sqrt{\frac{B\omega^{5}}{W\pi}}\,,   
\end{align}
\end{subequations}
where we have defined $t^*_{\text{even}}:=\tfrac{n\pi}{\omega}$ and $t^*_{\text{odd}}:=\tfrac{(2n+1)\pi}{2\omega}$. Due to the periodicity of $\rho_{HO}$, we can consider $t\in [0,\tfrac{2\pi}{\omega})$, then, the critical points reduces to $t^*_{\text{even}}=0$  and $t^*_{\text{odd}}=\tfrac{\pi}{2\omega}$. Notice that the second-time derivatives of $\rho_{HO}$ given by \eqref{secondderivative} are positive or negative quantities depending on whether $B$ is greater than or less than one. Furthermore, it is straightforward to see that for $B_> (B>1)$ (respectively $B_< (B<1) $ ) the critical points at $t^*_{\text{even}}$ are maximums (respectively minimums), while the critical points at $t^*_{\text{odd}}$ are minimums (respectively maximums). The values of $\rho$ at the critical points are 
\begin{subequations}
    \begin{align}
    \rho_{HO}(0,t^*_{\text{even}};B)=\left(\frac{B \omega}{\pi^{1/2} W} \right)^{1/2}\,,\\
    \rho_{HO}(0,t^*_{\text{odd}};B)=\left(\frac{\omega}{\pi^{1/2} B W} \right)^{1/2}\,.
    \end{align}
\end{subequations}

\begin{figure}[H]
    \centering
    \includegraphics[width=0.55\linewidth]{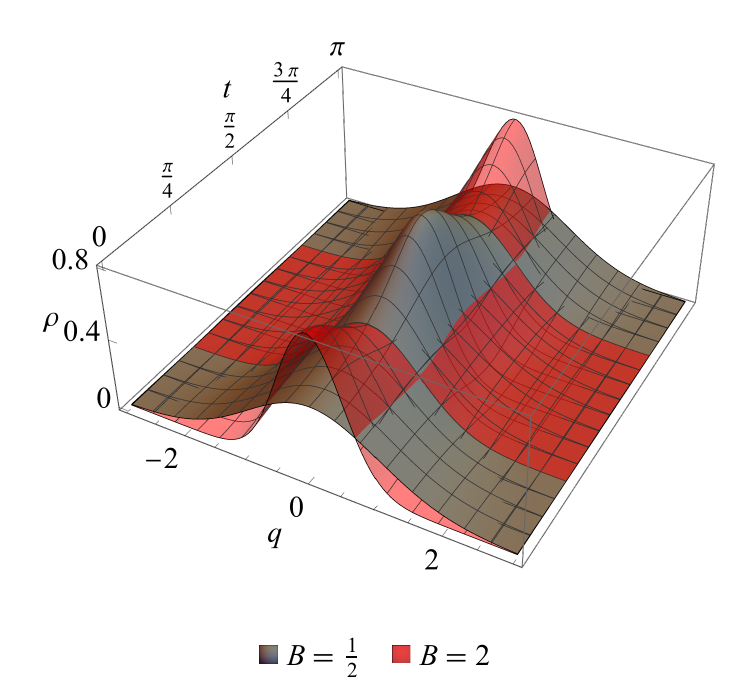}
    \captionsetup{font=small} 
    \caption{\justifying Distribution density probability  as a function of $q$ and $t$, with $X=W=1$ and $B=1/2,2$. With this choice of parameters, $\rho(q,t;1/2)$ and $\rho(q,t;2)$ crosses at $\tau=\pi/4,3\pi/4$ }
    \label{densidadprob}
\end{figure}

On the other hand, by the mean value theorem, there is a time $\tau\in(t^*_{\text{even}},t^*_{\text{odd}})$  such that $\rho_{HO}(0,\tau;B_>)=\rho_{HO}(0,\tau;B_<)$. In particular, for $B_<=B_>^{-1}$, it can be shown that \break $\rho_{HO}(0,\tau;B_>)=\rho_{HO}(0,\tau;B_>^{-1})$  at $\tau=\tfrac{\pi(2n+1)}{4\omega}$.  The previous analysis can be observed in Fig.~\ref{densidadprob}, where we have set $X=W=1$, and $B=1/2,2$.

\begin{figure}[H]
    \centering
    \includegraphics[width=0.55\linewidth]{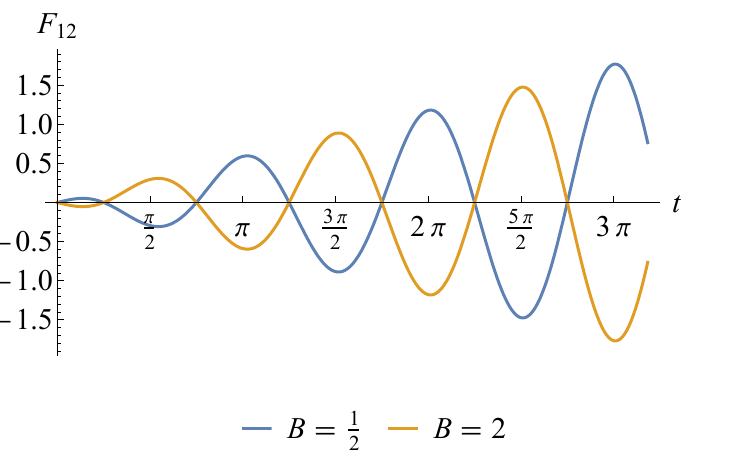}
    \captionsetup{font=small} 
    \caption{\justifying Components $F_{12}$ with $B=2$  and $B=1/2$, fixing $X=W=1$.}
    \label{berryo3}
\end{figure}

In Fig. \ref{berryo3} we plot the $F_{12}$ component of the tBerry curvature for $B=2$ and $B=1/2$. As we can see, their behavior is similar to that of the distribution density probability, with the maxima and minima at the same points. Furthermore, they increase over time, which is an advantage because it could be easier to measure experimentally.

\subsection{Region 2: inverted harmonic oscillator}

The quantum dynamics of the inverted oscillator have been investigated for several years \cite{Barton}. Over time, new characteristics of the system have been found, such as its relation to the zeros of the Riemann zeta function \cite {Osinv01}, or that it can be confined to give rise to a discrete spectrum \cite{Yuce}, and more recently its application in understanding the quantum mechanics of dispersion and time decay in various systems \cite{Osinv02}. Additionally, recent studies have shown that this system serves as an ideal toy model for understanding some ideas about chaos and complexity \cite{Osinv1, Osinv2, Osinv3}.
In this Section, we study the tQGT associated with the inverted oscillator by using a wave packet very similar to that described by Eq.~\eqref{eq:funciononda} for the harmonic oscillator. 

As we mentioned, we start by considering the initial state \eqref{inicial}. In this case, the kernel of the inverted oscillator is 
\begin{align}
    K(q',t;q,0)= \sqrt{\frac{\alpha}{2\pi i W t}}\exp{\left(-\frac{\alpha((q^2+q'^2)\cosh\left(\alpha t\right)-2qq')}{2iW\sinh\left(\alpha t\right)}\right)}\,,
    \label{kernelinv}
\end{align}
where $\alpha=\sqrt{-XW}$, which is a real parameter function since $X<0$ in this region.
Using this kernel, the evolution over time of the initial condition \eqref{inicial} results in
\begin{align}
\psi_{IHO}(q,t;\lambda)=&\left( \frac{B \alpha}{W \pi} \right)^{1/4} \frac{1}{\sqrt{\cosh \left(\alpha t\right)+i
   B \sinh \left(\alpha t\right)}}  \exp \left(- \frac{q^2}{2}\frac{ \alpha\left(1+i B \coth
  \left(\alpha t\right)\right)}{W \left(-B+i \coth \left(\alpha t\right)\right)}\right)\,.
   \label{eq:funcionondainvertido}
\end{align}
Comparing \eqref{eq:funcionondainvertido} and \eqref{eq:funcionondagen}, we obtain the identifications  
\begin{subequations}
\begin{align}
    U_{IHO}(t,\lambda)&=\frac{\alpha  B \operatorname{csch}^2\left(\alpha t\right)}{W \left(B^2+\coth^2 \left(\alpha t\right)\right)}\,,\\
    V_{IHO}(t,\lambda)&=-\frac{\alpha  \left(B^2+1\right) \coth \left(\alpha t\right)}{W \left(B^2+\coth^2\left(\alpha t\right)\right)}\,,\\
    C_{IHO}(t;\lambda)&=\left( \frac{B \alpha}{W \pi} \right)^{1/4} \frac{1}{\sqrt{\cosh \left(\alpha t\right)+i
   B \sinh \left(\alpha t\right)}}\,.
\end{align}
\end{subequations}

The expectation value of the Hamiltonian \eqref{firstHamil} on the state \eqref{eq:funcionondainvertido} turns out to be
\begin{align}
\bar E =\langle \hat{H} \rangle =\frac{(-1+B^2)\alpha}{4B}\,.
\label{valoresperadoII}
\end{align}
In this case, we also have two values of $B$ that solve \eqref{valoresperadoII} for the same average of the Hamiltonian $\bar E$, namely

\begin{align}
  B_1 = \frac{2\bar E +\sqrt{\alpha^2 +4 \bar E^2}}{\alpha} \,, \  \  B_2 = \frac{2\bar E -\sqrt{\alpha^2 +4 \bar E^2}}{\alpha}\,.
\end{align}
Notice that, only $B_1$ is physical, since $B_2$ is negative and then gives rise to a non-normalizable wave function.

Using the same varying parameters as before, $\Lambda=\{\Lambda^I\}= (t, X, W, B)$ with $I=0,1,2,3$, the components of the tQMT are given by
\begin{subequations} \allowdisplaybreaks
\label{metricahiperbolica}
    \begin{align}
    g_{00}&=\frac{\left(B^2+1\right)^2\alpha^2}{8 B^2}\,,\\
    g_{01}&=-\frac{\left(B^2+1\right) W \left(-\left(B^2-1\right) \sinh \left(2 \alpha t\right)+2 \left(B^2+1\right) \alpha t\right)}{32 B^2 \alpha}\,,\\
   g_{02}&-\frac{\left(B^2+1\right) X \left(\left(B^2-1\right) \sinh \left(2 \alpha  t\right)+2 \left(B^2+1\right) \alpha t \right)}{32 B^2 \alpha}\,,\\
   g_{03}&=0 \,,\\
    g_{11} &=\frac{-8 \left(B^4-1\right)  \alpha  t \sinh \left(2  \alpha t\right)-B^4+8
   \left(B^2+1\right)^2  \left(\alpha t \right)^2+\left(B^2+1\right)^2 \cosh \left(4  \alpha t\right)+6 B^2-1}{256 B^2 X^2}\,, \\
     g_{12} &=-\frac{B^4+8 \left(B^2+1\right)^2 \left(\alpha t\right)^2 -\left(B^2+1\right)^2 \cosh \left(4 t
  \alpha\right)-6 B^2+1}{256 (B\alpha)^2 }\,, \\ 
   g_{22} &=\frac{8 \left(B^4-1\right) \alpha t \sinh \left(2 \alpha t\right)-B^4+8
   \left(B^2+1\right)^2 \left(\alpha t\right)^2+\left(B^2+1\right)^2 \cosh \left(4 \alpha t\right)+6 B^2-1}{256 B^2 X^2}\,, \\
    g_{13} &=\frac{\cosh \left(2 \alpha t\right)}{16 B X}\,,\\
    g_{23} &=-\frac{\cosh \left(2 \alpha t\right)}{16 B W}\,,\\
     g_{33} &=\frac{1}{8 B^2}\,.
    \end{align}
\end{subequations}    

From \eqref{metricahiperbolica}, we can see that some components exhibit a hyperbolic function-like time dependence, leading to an exponential asymptotic behavior as $\alpha t \to \infty$. We now analyze the behavior of these metric components, paying particular attention to the asymptotic limit as $\alpha t \to\infty$. Here, we can distinguish three different cases. 

\begin{figure}[H]
 \centering
  \begin{subfigure}{0.32\linewidth}
    \includegraphics[width=\textwidth]{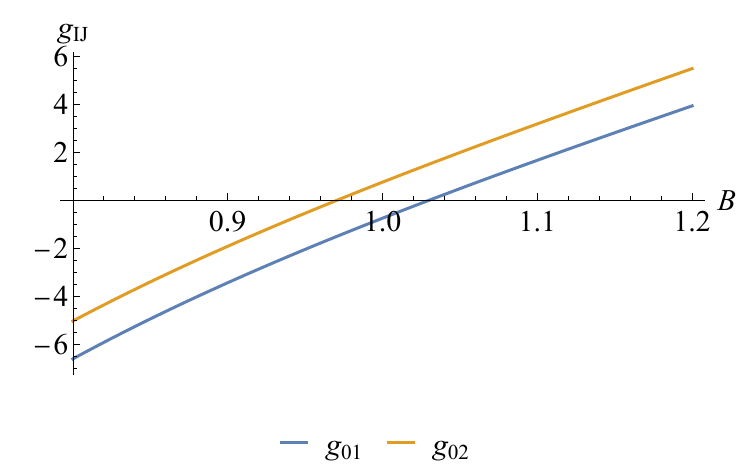}
    \caption{}
    \label{fig:gvariandobc}
  \end{subfigure}
  \centering
  \begin{subfigure}{0.32\linewidth}
    \includegraphics[width=\textwidth]{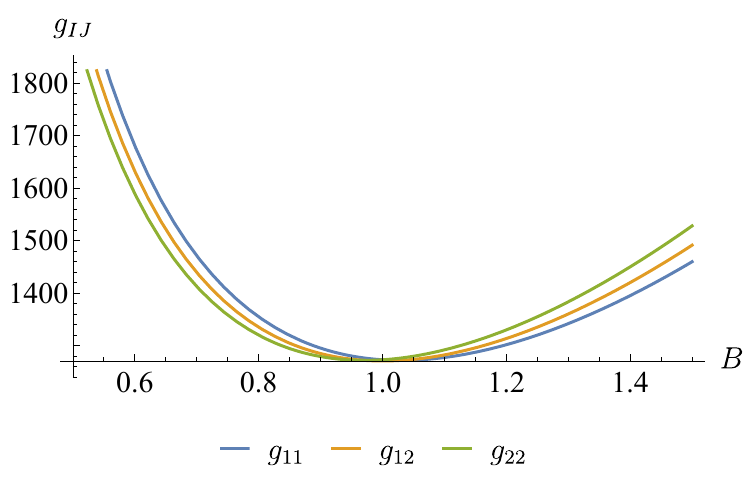}
    \caption{}
    \label{fig:gvariandoba}
  \end{subfigure}
  \begin{subfigure}{0.32\linewidth}
    \includegraphics[width=\textwidth]{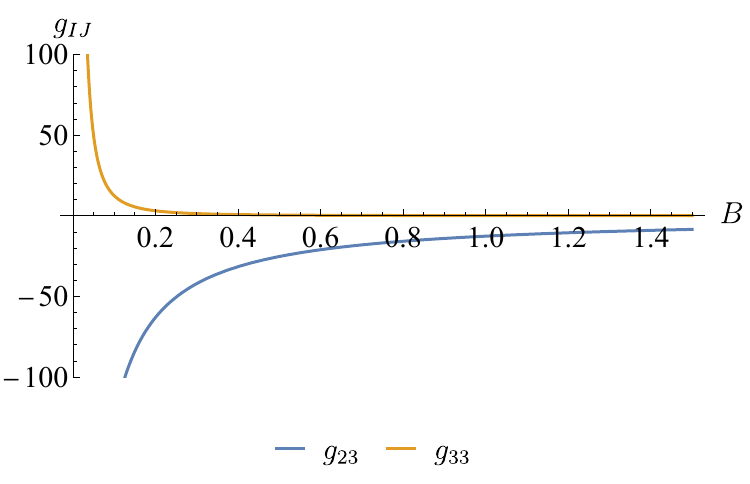}  
    \caption{}
    \label{fig:gvariandobb}  
  \end{subfigure}
  \captionsetup{font=small} 
  \caption{\justifying  Components of the tQMT for the region 2, as a function of the parameter $B$, all of them are singular at $B=0$. (a) Components $g_{01}$ and $g_{02}$ of the tQMT. $g_{01}$ crosses zero at the right of $B=1$, while $g_{02}$ crosses zero at the left of $B=1$. (b) These components have a minimum around $B=1$. (c) These components show a monotone behavior.}
\label{fig:gvariandobtodas}
\end{figure}

\begin{enumerate}
 \item The components $g_{01}$ and $g_{02}$ lose their hyperbolic time dependence at $B=1$. Furthermore, $g_{01}$ and $g_{02}$ vanish at $B_{I+}$ and $B_{I-}$, respectively, with
    \begin{align} 
    \label{BI}
    B_{I\pm}&=\sqrt{\frac{\sinh\left(2 \alpha t\right)\pm 2\alpha t} {\sinh\left(2 \alpha t\right)\mp 2 \alpha t}}\,.
\end{align}
This means that $g_{01}$ crosses zero at $B<1$, while $g_{02}$ crosses zero at $B>1$. Notice that $B_{I\pm}\to 1$ when $\omega t\to \infty$. Now, we proceed similarly to the harmonic case, rewriting \eqref{BI} as $ B_{I\pm}=1\pm \Delta B_I$. Hence, `the neighbourhood of $B=1$' in this context means the interval $[ 1-|\Delta B_I|,1+|\Delta B_I| ]$, with $\Delta B_I\ll 1$. Neglecting second-order terms on $\Delta B_I$, we find  $\Delta B_I \approx \tfrac{ 2\alpha t}{\sinh 2\alpha t}$. Then, for large $\alpha t$ (i.e. $\alpha t$ such that $\tfrac{ 2\alpha t}{\sinh 2\alpha t} \,\ll 1$), the components $g_{01}$ and $g_{02}$ are equal to zero in the neighborhood of 1, which corresponds to the point where the average energy changes from positive to negative.  This behavior can be appreciated in Fig. \ref{fig:gvariandobc}, where we have set $X=-1,$ $W=1$, and $t=3$.  For this choice of parameters, we have $\Delta B \approx 0.0297.$
  \item  The component $g_{00}$ have a minimum at $B=1$. Furthermore, $g_{11}$, $g_{22}$, and $g_{12}$ have a minimum at $B_{I+}$, $B_{I-}$, and $B=1$ respectively. Then, for the larger $ \alpha t$, the minimum of $g_{11}$ and $g_{22}$ is closer to $B=1$. In particular, this minimum occurs at $B\to 1$ as $\alpha t \to \infty$. See  Fig. \ref{fig:gvariandoba}
    
   \item  The components $g_{13}$, $g_{23}$, and $g_{33}$ have a monotonic behavior. See Fig. \ref{fig:gvariandobb}, where we have used $X=-1,$ $W=1$, and $t=3$, for which $g_{13}=g_{23}.$
\end{enumerate}

\begin{figure}[H]
    \centering
{\includegraphics[width=0.55\linewidth]{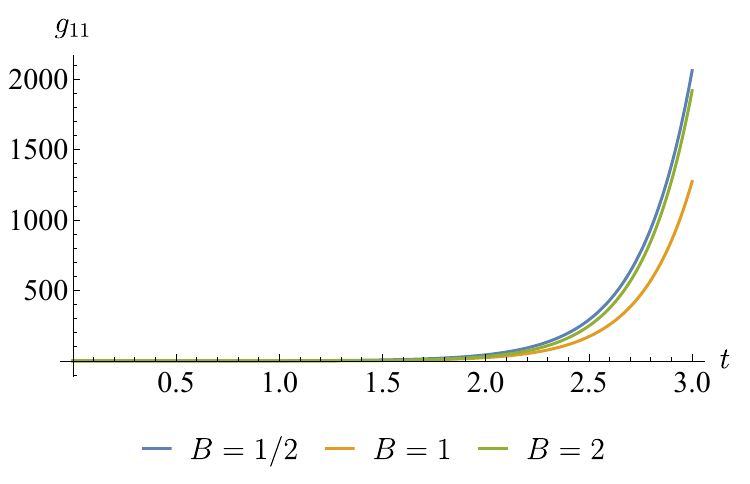}}
    \captionsetup{font=small} 
    \caption{Component $g_{11}$ of the tQMT in the region 2 as a function of time. The parameters were fixed as $X=-1$ and $W=1$.}
\label{fig:g11variandot}
\end{figure}

In Fig.~\ref{fig:g11variandot} we plot $g_{11}$ as a function of $t$. Here we can see that this metric component exhibits an exponential behavior for large $t$. The other metric components with time dependence have a similar behavior. 

Now we will focus on tBerry curvature. The components of this tensor are
\begin{subequations}
    \begin{align}
      F_{01}&=-\frac{\left(B^2+1\right) W \cosh \left(2  \alpha t \right)}{8 B \alpha}\,,\\      
      F_{02}&=\frac{\left(B^2+1\right) X \cosh \left(2 \alpha t \right)}{8 B \alpha} \,,\\     
      F_{03}&=\frac{\left(B^2+1\right) \alpha}{4 B^2}\, ,\\
      F_{12}&=\frac{\left(B^2+1\right) t \cosh \left(2  \alpha t \right)}{8 B \alpha}\,,\\
      F_{13}&=\frac{2 \left(B^2+1\right) t
   \alpha-\left(B^2-1\right) \sinh \left(2 \alpha t \right)}{16 B^2 X}\,,\\
      F_{23}&=\frac{2 \left(B^2+1\right) t
   \alpha +\left(B^2-1\right) \sinh \left(2 \alpha  t\right)}{16 B^2 W}\,.
    \end{align}
\end{subequations}

\begin{figure}[H]
  \centering
  \begin{subfigure}{0.45\linewidth}
    \includegraphics[width=\textwidth]{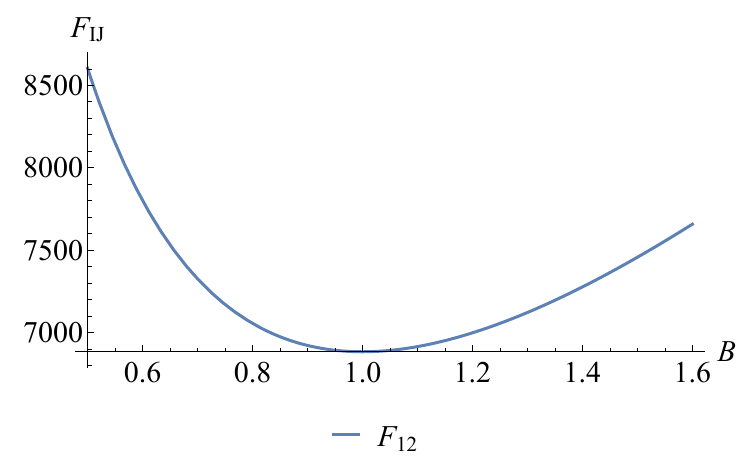}
    \caption{}
    \label{fig:fvariandoba}
  \end{subfigure}
  \begin{subfigure}{0.45\linewidth}
    \includegraphics[width=\textwidth]{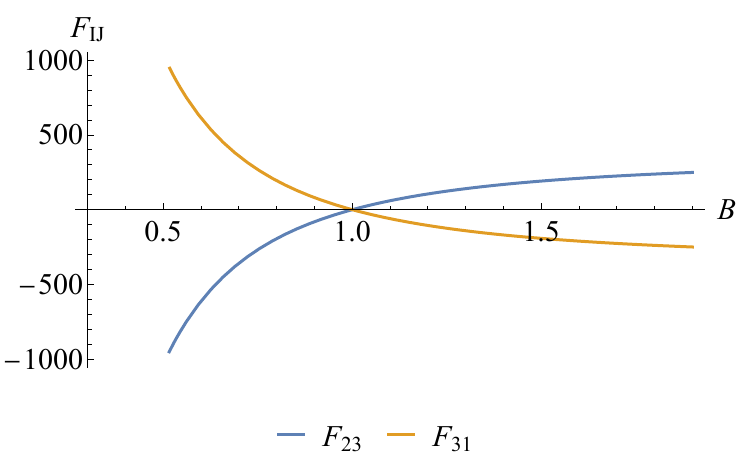}  
    \caption{}
    \label{fig:fvariandobb}  
  \end{subfigure}
  \captionsetup{font=small} 
  \caption{\justifying Components of the tBerry curvature for the region 2, as functions of the parameter $B$. The remaining parameters are fixed as follows: $X=-1,$ $W=1$, and $t=5$. (a) This component has a minimum at $B=1$, $F_{01}$ and $F_{02}$ have a similar behavior. (b) These components show a monotone-decreasing behavior.}
    \label{fig:fvariandob}
\end{figure}

\begin{figure}[H]
    \centering
    \includegraphics[width=0.55\linewidth]{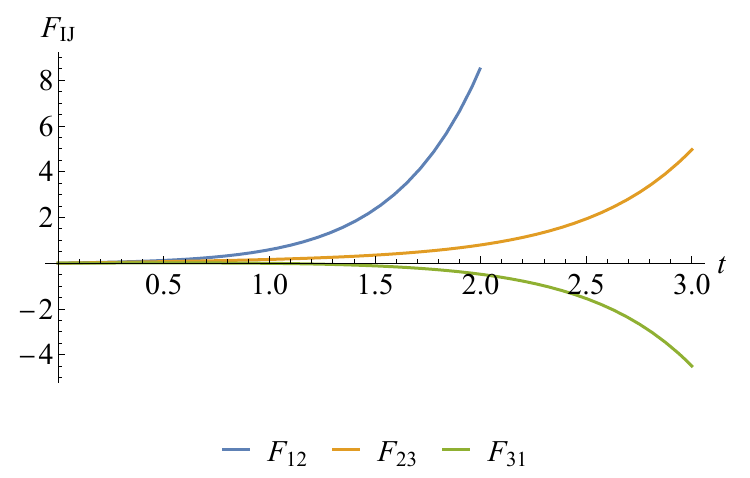}
  \captionsetup{font=small} 
    \caption{\justifying Components of the tBerry curvature in the region 2 as a function of time. The parameters were fixed as $W=B=1$ and $X=-1$.}
    \label{fig9}
\end{figure}

Let us first examine the behavior of these components as a function of $ B $.  Notice that at $B=1$ the components $ F_{01} $ and $ F_{02}$ exhibit a minimum, while the component $F_{12}$ has a maximum. Furthermore, the components $ F_{13} $ and $ F_{23}$ are zero at $B=B_{+}$ and $B=B_{-}$, respectively. The behavior of these components as a function of $B$ is illustrated in Fig. \ref{fig:fvariandob}. On the other hand, excepting $F_{03}$, all the components show a hyperbolic behavior as a function of $t$ (see Fig. \ref{fig9}).

\subsection{Transition zone}

\begin{figure}[H]
  \centering
  \begin{subfigure}{0.45\linewidth}
    \begin{tikzpicture}
    \node[anchor=south west,inner sep=0] (g11variandoX) at (0,0) {\includegraphics[width=0.9\linewidth]{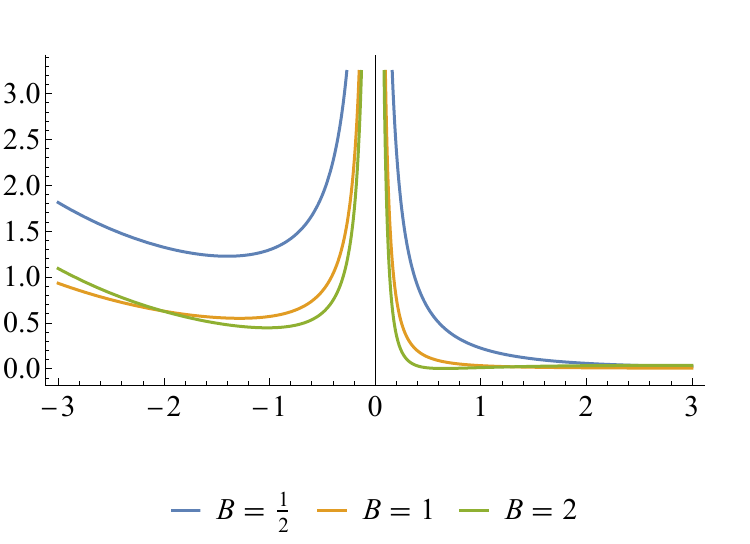}};
    \begin{scope}[x={(g11variandoX.south east)},y={(g11variandoX.north west)}]
        \node at (0.06,0.95) {{\scriptsize$g_{11}$}};
        \node at (0.98,0.3) {{\scriptsize$X$}};
    \end{scope}
\end{tikzpicture}
    \caption{}
    \label{fig:gvariandoXa}
  \end{subfigure}
  \begin{subfigure}{0.45\linewidth}\begin{tikzpicture}
    \node[anchor=south west,inner sep=0] (f31variandox) at (0,0) {\includegraphics[width=0.9\linewidth]{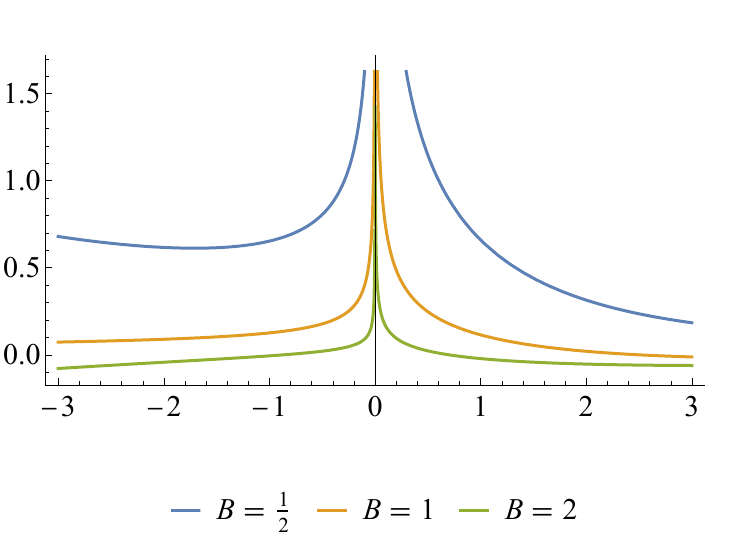}};
    \begin{scope}[x={(g11variandoX.south east)},y={(g11variandoX.north west)}]
        \node at (0.06,0.95) { {\scriptsize$F_{31}$}};
        \node at (0.98,0.3) {{\scriptsize $ X$}};
    \end{scope}
\end{tikzpicture}
    \caption{}
    \label{fig:fvariandoXb}  
  \end{subfigure}
  \captionsetup{font=small} 
  \caption{\justifying (a) Component $g_{11}$ of the tQMT, and (b) component $F_{31}$ of the tBerry curvature with $t=1$. Both $g_{11}$ and $F_{31}$ are singular $X\to 0$, corresponding to the bifurcation point from the harmonic oscillator to the inverted oscillator.}
    \label{fig:fvariandoX}
\end{figure}

The goal now is to analyze the behavior of the tQMT near the bifurcation point at $X=0$, where the system transitions from a harmonic to an inverted oscillator. The key features of the tQMT at the vicinity of $X=0$ are:

\begin{enumerate}
    \item The parameter component $g_{ij}$ in region 1 has the same asymptotic behavior of the corresponding $g_{ij}$ in region 2, at $X\to 0$.
    \item In both regions, the components $g_{11}$, $g_{12}$ and $g_{13}$ diverge at $X \to 0$,  their asymptotic behavior at this point is $g_{11} \sim \frac{1}{32X^2}$, $g_{12} \sim -\frac{1}{32XW}$ and $g_{13} \sim \frac{1}{16XB}$ In Fig. \ref{fig:gvariandoXa}  we plot $g_{11}$, showing its behavior at at $X \to 0$.
    \item The components $g_{00}$, $g_{01},$ and $g_{02}$ are continuous but not differentiable at $X=0$.
\end{enumerate}

Regarding the tBerry curvature, we observe a behavior analogous to that of the tQMT in the vicinity of $X=0$. At $X\to 0$, the components $F_{12}$ and $F_{13}$ are singular  (see Fig. \ref{fig:fvariandoXb}), while the remaining non-zero components are continuous but not differentiable.

The analysis clearly shows that the metric components $g_{1i}$ and tBerry curvature components $F_{1I}$ diverge at $X\to 0$, representing the point at which the system shifts from a Harmonic oscillator to an inverted one.

\section{Harmonic oscillator with time-dependent frequency}\label{secHOTDF}

In this section, we study a system whose Hamiltonian explicitly depends on time. We consider the harmonic oscillator with time-dependent frequency (HOTDF), whose Hamiltonian is given by 
\begin{align}
    \hat{H}(\hat{q},\hat{p},\Lambda)=\frac{\hat{p}^2}{2}+\omega(\Lambda)^2\frac{\hat{q}^2}{2}\,,
\end{align}
where $\Lambda=\{\Lambda^I\}$ ($I,J,\dots\!=0,1,\dots,\mathcal{N}$) is the set of time and parameters. In general, methods to solve problems with explicit time dependence are an open subject nowadays. An approach for solving these types of problems is to find constants of motion or invariants. 

For the HOTDF there exists an invariant operator $\hat{L}$, known as the Lewis invariant, which is given by
\begin{align}
    \hat{L}=\frac{1}{2} \left( \frac{\hat{q}^2}{\gamma^2}+(\gamma \hat{p} -\hat{q} \dot{\gamma})^2  \right)\,,
\end{align}
where $\gamma$ satisfies the Ermakov equation
\begin{align}
\frac{\mathrm{d}^2 \gamma}{\mathrm{d} t^2}+\omega^2(\Lambda)\gamma -\frac{1}{\gamma^3}=0\,.
\end{align}
The existence of this invariant operator is of utmost importance in solving the corresponding time-dependent Schrodinger equation. Lewis and Riesenfeld \cite{Lewis,Dittrich} have shown that the general solution is given by the superposition
\begin{align}  \ket{\Psi(\Lambda)}=\sum_n c_n e^{i \beta_n(t)} \ket{n_L(\Lambda)}\,,
\label{superposition}
\end{align}
where $\ket{n_L(\Lambda)}$ are the normalized eigenkets  of the Lewis operator, i.e.,
\begin{align}
    \hat{L}\ket{n_L(\Lambda)}=\left(n+\tfrac{1}{2} \right)\ket{n_L(\Lambda)}\,,
    \label{lewiseigen}
\end{align}
and $\beta_n(t)$ are time-dependent phases, obtained from the equation 
\begin{align}
    \frac{\mathrm{d}\beta_n}{\mathrm{d}t}=  \dot{\beta}_n= \bra{n_L(\Lambda)} i\partial_t-\hat{H} \ket{n_L(\Lambda)}\,.
    \label{fasebeta}
\end{align}

The solutions to equations \eqref{lewiseigen} and \eqref{fasebeta}  are, respectively, given by
\begin{subequations}
\begin{align}
\psi_{n_L}(q;\Lambda)&=\braket{q}{n_L(\Lambda)} =\frac{\exp \left(-\left(\frac{1}{\gamma^2}-i \frac{\dot{\gamma}}{\gamma} \right)\frac{q^2}{2}\right)}{\sqrt{2^n n! |\gamma| \sqrt{\pi} }} H_n\left( \frac{q}{|\gamma|}\right)\,,
\label{eigenLewis} \\
\beta_n(t)&=-\left(n+\frac{1}{2}\right)h(t) \,, 
\end{align}
\end{subequations}
with $h$ obtained through the equation $\dot{h}=\tfrac{1}{\gamma^2}$. To simplify our calculations, we consider the case where all the coefficients $c_n$ involved in \eqref{superposition}, except one, are zero. Hence, under this consideration \eqref{superposition} reduces to
\begin{align}  \label{ketLewis}
\ket{\Psi(\Lambda)}= e^{i \beta_n(t)} \ket{n_L(\Lambda)}\,.
\end{align}

We now proceed to compute the tQMT associated with the quantum state \eqref{ketLewis}. Analogously to what happens in the case of a stationary state, the tQMT turns out to be independent of the time-dependent phase $\beta_n(t).$ In fact, substituting \eqref{ketLewis} into \eqref{QGT-time} we get
\begin{align}\label{QLewis}
      Q_{IJ}(\Lambda)&=\braket{\partial_I n_L}{\partial_J n_L} -\braket{\partial_I n_L}{n_L}\braket{n_L}{\partial_J n_L}\,.
\end{align}
However, in contrast to the stationary state case (see \eqref{Qstationary}), \eqref{QLewis} is time-dependent since $\ket{n_L(\Lambda)}$ is.

Using the eigenfunction \eqref{eigenLewis}, and taking the real and imaginary part of \eqref{QLewis}, we find that the components of the tQMT and the tBerry curvature are
\begin{subequations}
\begin{align}
    g_{IJ}  
    &=\frac{\gamma^4(n^2+n+1)}{8}\left( \partial_I \left( \frac{\dot{\gamma}}{\gamma}\right)\partial_J \left(\frac{\dot{\gamma}}{\gamma}\right)  +\left(\partial_I\gamma^{-2}\right)\left(\partial_J\gamma^{-2}\right)\right)\,,\label{metricatiempog}\\
     F_{IJ}  
    &=\frac{\gamma^4(2n+1)}{4}\left(\left(\partial_J \gamma^{-2}\right)\partial_I \left( \frac{\dot{\gamma}}{\gamma}\right)-\left(\partial_I\gamma^{-2}\right)\partial_J \left( \frac{\dot{\gamma}}{\gamma}\right)\right)\,, \label{eq:BCurvtiempog}
\end{align}
\end{subequations}
respectively.

This system exhibits some properties analogous to those of the model described by the wavefunction \eqref{eq:funcionondagen}. In particular, it can be verified that the metric \eqref{metricatiempog} satisfies the relations \eqref{eq:determinante0} and \eqref{eq:curvatura1d}. Nevertheless, instead of \eqref{eq:palumbo}, the relation between \eqref{metricatiempog} and \eqref{eq:BCurvtiempog} is given by
\begin{align}
    \det\tilde{ \boldsymbol{g}}[ \Lambda^{\scriptscriptstyle I_1},\Lambda^{ \scriptscriptstyle I_2}]&= \frac{(2n+1)^2}{4 (n^2+n+1)^2}\left(F_{\scriptscriptstyle I_1 I_2} \right)^2 \,, \qquad \mathrm{if} \, \, \mathcal{N}=2\,, \label{eq:Palumbdomodi}
\end{align}
which coincides with \eqref{eq:palumbo} for $n=0$.

Let us now consider a specific example in which the frequency is given by
\begin{align}
\omega(t)=\sqrt{\omega_0^2e^{2At}-\frac{A^2}{4}}\,, \label{eq:partfre}
\end{align}
where $\omega_0^2e^{2At}-\frac{A^2}{4}>0$. Then, the set of time and system parameters is $\Lambda=\{\Lambda^I\}=\{ t, A, \omega_0 \}$, with $I=0,1,2$. Notice if $A=0$, the frequency becomes constant, recovering the usual harmonic oscillator with squared frequency $\omega^2_0$. For the frequency \eqref{eq:partfre} we obtain the explicit solution for $\gamma$ and $h$, obtaining $\gamma=e^{-At/2}$ and $h=\omega_0 e^{At}/A$. Substituting these expressions into \eqref{metricatiempog} and \eqref{eq:BCurvtiempog}, we find that the  metric is
\begin{equation}
g_{IJ}=\frac{n^2+n+1}{8}\begin{pmatrix}
A^2 & At & \frac{A}{\omega_0}\\
At & \frac{e^{-2At}}{4\omega_0^2}+t^2 & \frac{t}{\omega_0}\\
 \frac{A}{\omega_0} & \frac{t}{\omega_0} & \frac{1}{\omega_0^2}
\end{pmatrix}\,,
\end{equation}
and that the tBerry curvature is
\begin{equation}
F_{IJ}=\frac{(2n+1)e^{-At} }{8\omega_0^2}\begin{pmatrix}
0 & A\omega_0 & 0\\
- A\omega_0  & 0 & -1\\
0 &   1& 0
\end{pmatrix}\,. \label{eq:curt}
\end{equation}

Notice that the component $g_{00}$ is time-independent, implying that the energy dispersion remains constant in this case. Also, components $g_{0I}$ and $F_{0I}$ vanish if $A=0$, in which case our state becomes a stationary state. Furthermore, for  $t \to \infty$ the tBerry \eqref{eq:curt} tends to $0$, corresponding to a zero phase interference in this limit.

Notice that the system corresponds to a free particle in the limit $\omega (t)=0$. There are two cases where frequency \eqref{eq:partfre} vanishes: a) both parameters $A$ and $\omega_0$ tend to zero and b)  $4\omega_0^2-A^2 e^{-2At}$ tends to zero. Both cases are reflected in the metric and tBerry curvature. 
For case a) we have that the metric components $g_{00}$ and $g_{01}$ tend zero and the rest become undefined. For the curvature, all its components become undefined. Then, as expected, we have a breakdown of the geometric structure in this limit because the system would correspond to a free particle and not to an oscillator anymore.
On the other hand, for b) we have the curvature component $F_{01}$ becomes constant, i.e., independent of the parameters. This is reflected in the metric, because of the Palumbo-Goldman relation \eqref{eq:Palumbdomodi},  as $ \det\tilde{ \boldsymbol{g}} [t, A] \to  \frac{n^2+n+1}{8}$.


\section{Coupled systems}\label{sec:coupledsystems}

In this Section, we consider a system of $N$ coupled particles described by a wave function corresponding to a generalization of \eqref{eq:funcionondagen} and analyze the associated tQMT and tBerry curvature. In normal coordinates $Q=\{ Q_a \}$ with $a=1, \dots, N$, the generalized wave function is given by
\begin{align}
\psi(Q,\Lambda)=C\left(\Lambda\right)\exp{\left(-\sum_{a=1}^N \left(U^a(\Lambda) +iV^a(\Lambda) \right) Q_a^2/2 \right)}\,,
\label{eq:funcionondagenacoplado}
\end{align}
where $U^a(\Lambda) > 0$, $V^a(\Lambda)\in \mathbb{R}$, $C(\Lambda)\in \mathbb{C}$, and $\Lambda=\{\Lambda^I\}$ ($I,J,\dots\!=0,1,\dots,\mathcal{N}$) is the set of time and parameters. The normalization condition of the wave function implies
\begin{align}
|C(\Lambda)|^2=\pi^{-N/2}\prod_{a=1}^N \sqrt{U^a(\Lambda)}\,.
\end{align}

Using~\eqref{metric-tensor-time} and \eqref{eq:funcionondagenacoplado}, it is found that the components of the tQMT are given by
\begin{align}\label{metric_coupled}
g_{IJ}=\sum_{a=1}^N\frac{\partial_{I}U^{a}\partial_{J}U^{a}+\partial_{I}V^{a}\partial_{J}V^{a}}{8(U^{a})^2}\,.
\end{align}
Notice that the tQMT has dimensions of $( \mathcal{N}+1 )\times ( \mathcal{N}+1 )$. On the other hand, if we \break consider that all functions $U^a$ and $V^b$ are distinct from each other and that there is no \break functional dependency between them (i.e., $U^b\neq f(U^1,..., U^{b-1}, U^{b+1},\dots, U^N, V)$ and \break $V^b\neq f(U, V^1,..., V^{b-1}, V^{b+1},\dots, V^N)$), then the metric is generated by a set of $2N$ independent functions $\{ U^a, V^a \}$. Here $U= \{ U^a\}$ and $V= \{ V^a\}$. It can be shown by induction on $N$ that the determinant of the metric is zero if $\mathcal{N}=2N$. 

Now, let us consider the case where there is functional dependency among the functions $U$ and $V$ with multiplicity $l$. Then, we can define a constraint surface given by $l$ functions that relate the functions $U^a$ and $V^a$ among themselves, this can be written as
\begin{align}
    f_1(U,V)&=0\,,\nonumber\\
    &\vdots\nonumber \\
    f_{l}(U,V)&=0 \,.
\end{align}

In this scenario, the determinant of the metric \eqref{metric_coupled} is zero if $\mathcal{N}>2N-l-1$, which means that there can be at most $2N-l$ independent parameters. Taking this into account, in the following, we consider metrics $\tilde{\boldsymbol{g}}[\Lambda^{\scriptscriptstyle I_1},\dots,\Lambda^{\scriptscriptstyle I_{\eta}}]$ obtained from the variation of $\eta$ elements $\Lambda^{\scriptscriptstyle I_1},\dots,\Lambda^{I_{\scriptscriptstyle \eta}}$ (with $I_1,\dots, I_{\eta}\in { 0,\dots, \mathcal{N} }$). 

For a more comprehensive analysis, it would be of interest to calculate the scalar curvature $\mathcal{R} [\Lambda^{I_1},\dots,\Lambda^{I_{\eta}}]$ associated with the metrics $\tilde{\boldsymbol{g}}[\Lambda^{\scriptscriptstyle I_1},\dots,\Lambda^{\scriptscriptstyle I_{\eta}}]$. Nevertheless, the calculations for the scalar curvature are quite involved when the functions $U^a$ and $V^a$ are arbitrary, and hence, we restrict ourselves to the case $N=2$, where the maximum number of independent parameters is $\eta=4$. In addition, the resulting expressions for the Ricci tensor for $\eta=2,3$, when keeping $U^a$ and $V^a$ arbitrary, are too lengthy and do not yield substantial information. In the case $N=2$ and $\eta=4$, for arbitrary functions $U^a$ and $V^a$, the scalar curvature takes the remarkable simple form 
\begin{equation}
\mathcal{R}[\Lambda^{I_1},\dots,\Lambda^{I_{4}}]=-32\,.
\end{equation}

This expression and \eqref{eq:curvatura1d}, which corresponds to the case where $N=1$ and $\eta=2$, suggest the possibility of obtaining an analogous expression for the scalar curvature $\mathcal{R}[\Lambda^{I_1},\dots,\Lambda^{I_{2N}}]$ in the case $N>2$ and $\eta=2N$. However, the calculations for $N>2$, using arbitrary functions $U^a$ and $V^a$, result in extensive procedures. Particularly, for several specific examples and $N=3,4,5,6$, we find that the scalar curvature is given by 
\begin{align}
\label{escalar16n}
    \mathcal{R}[\Lambda^{I_1},\dots,\Lambda^{I_{2N}}]=-16N\,.
\end{align}
Although this is not a formal proof, it suggests the hypothesis that this property holds for all $N$ and any set of functionally independent functions $U^a$ and $V^a$.

On the other hand, using~\eqref{Berry-tensor-time} and \eqref{eq:funcionondagenacoplado}, we find that the components of the tBerry curvature are
\begin{align}
F_{IJ}=\sum_{a=1}^N \frac{-\partial_{I}U^a\partial_{J}V^a+\partial_{I}V^a\partial_{J}U^a}{4(U^{a})^2} \,.
\end{align}

The previous cases will now be analyzed in detail for a chain of generalized oscillators, where the functions $U^a$ and $V^a$ are known. Specifically, a system that admits solutions of the form \eqref{eq:funcionondagenacoplado} is a generalization of the system of one particle described by Hamiltonian \eqref{firstHamil} to a system of $N$ particles described by the Hamiltonian
\begin{align} 
\label{hamiltonianoqpyfija}
\hat{H}( \hat{q},\hat{p}, \Lambda)=\dfrac{1}{2} \hat{p}^{\intercal} \hat{p}+\dfrac{1}{2} \hat{q}^{\intercal} \boldsymbol{K} \hat{q} +\frac{Y}{2} \left( \hat{q}^{\intercal} \hat{p} +\hat{p}^{\intercal} \hat{q} \right)\,,
\end{align}
where $\hat{q}=(\hat{q_1},\dots,\hat{q}_N)^{\intercal}$, $\hat{p}=(\hat{p}_1,\dots,\hat{p}_N)^{\intercal}$, and
\begin{align}
\boldsymbol{K}= Z \boldsymbol{A}+X\boldsymbol{1} \,, 
\end{align}  
with $\boldsymbol{A}$ a $N \times N$ matrix of constant coefficients and $\boldsymbol{1}$ the $N \times N$ identity matrix. It is worth noting that the Hamiltonian \eqref{hamiltonianoqpyfija} involves the term with $Y$, which is not present in Hamiltonian \eqref{firstHamil}. Because of this, in one dimension, this system is referred to as the generalized harmonic oscillator and it is well-known to exhibit non-trivial Berry curvature \cite{chruscinski}.

Through the canonical transformation $\hat{q}=\boldsymbol{S}\hat{Q}$ and $\hat{p}=\boldsymbol{S} \left( \hat{P}-Y \hat{Q} \right)$, the Hamiltonian \eqref{hamiltonianoqpyfija} can be expressed as
\begin{align}
\hat{H}(\hat{Q},\hat{P},  \Lambda)=\dfrac{1}{2}\hat{P}^{\intercal}\hat{P}+\dfrac{1}{2}\hat{Q}^{\intercal}\boldsymbol{\Omega}^2\hat{Q}\,.
\end{align}
Here, the matrix $\boldsymbol{S}$ diagonalizes $\boldsymbol{M}=\boldsymbol{K}-Y^2\boldsymbol{1}$, and $\boldsymbol{\Omega}=\operatorname{diag}(\omega_1, \dots, \omega_N)$, where $\omega_a^2$, with $a=1, \dots, N$, represents the eigenvalues of $\boldsymbol{M}$; that is, $\boldsymbol{S}^{\intercal} \boldsymbol{M} \boldsymbol{S}=\boldsymbol{\Omega}^2$. The normal frequencies $\omega_b$ are given by
\begin{align}
    \omega_b=\sqrt{a_b Z+X-Y^2}\,,
\end{align} 
where $a_b$ are the eigenvalues of $\boldsymbol{A}$.

We assume that $\omega_{c+1}^2>\omega_{c}^2$ for $c=1,\dots, N-1$. This condition requires that all eigenvalues of $\boldsymbol{A}$ be distinct. Analogous to the case $N=1$, where we have considered two regions, in the general $N$-dimensional system we can consider $N+1$ regions, depending on the value of the parameters. The $m$-th region, with $m=1,..., N+1$, corresponds to the first $m-1$ normal modes behaving as inverted harmonic oscillators, while the remaining $N-m+1$ modes are harmonic oscillators. For $a_b Z+X-Y^2<0$, we introduce the real parameter functions $\alpha_b=\sqrt{Y^2-a_bZ-X}$ $(b=1,\dots,N)$. In this scenario, it is convenient to rewrite the Hamiltonian for the $m$-th region with $m=2,\ldots, N$ as
\begin{align}\label{Hamilregm}
    H_{m}(\hat{Q},\hat{P},  \Lambda)=&\frac{1}{2}\sum_{c=1}^{m-1} ( \hat{P}_c^2-\alpha_c^2 \hat{Q}_c^2)+\frac{1}{2}\sum_{d=m}^{N} (\hat{P}_d^2+\omega_d^2 \hat{Q}_d^2)\,,
\end{align}
where we have introduced the sub-index $m$ to point out that we are in the  $m$-th region (alternative, that we have $m-1$ normal modes that behave as inverted oscillators, which we call {\it{inverted normal modes}}). In the case $m=1$ there are only harmonic normal modes, whereas in the case $m=N+1$ there are exclusively inverted normal modes.

As in the previous Section, we consider solutions of the time-dependent Schr\"{o}dinger equation for the Hamiltonian \eqref{Hamilregm} that are non-time-separable. More precisely, we consider solutions of the form
\begin{align}
\Psi_{m}(Q,\Lambda)= \prod_{d=1}^{m-1}\chi_d(Q_d,\Lambda) \prod_{c=m}^{N}\psi_c(Q_c,\Lambda) \,,
\label{eq:funciondeondaacoplados}
\end{align}
where
\begin{align}\label{solIH}
\chi_d (Q_d,\Lambda) = \left(\frac{B_d \alpha_d }{\pi}\right)^{1/4} \frac{ \exp \left(-\frac{Q_d^2 \alpha_d}{2} \frac{\left(1+i B_d \coth \left(t
  \alpha_d\right)\right)}{ \left(-B_d+i  \coth \left( \alpha_d t\right)\right)}-\frac{i Y Q_d^2}{2} 
   \right)}{\sqrt{ \cosh \left(\alpha_d t\right)+i B_d \sinh \left( \alpha_d t \right)}}\,,
\end{align}
is a solution for an inverted normal mode and
\begin{align}
   \psi_c(Q_c,\Lambda)= \left( \frac{B_c \omega_c}{\pi} \right)^{1/4} \frac{\exp \left(- \frac{Q_c^2\omega_c}{2}\frac{ \left(1-i B_c \cot
   \left(\omega_c t\right)\right)}{ \left(B_c-i \cot \left( \omega_c t \right)\right)}-\frac{i Y Q_c^2}{2} \right)}{\sqrt{\cos \left( \omega_c t\right)+i
   B_c \sin \left( \omega_c t \right)}}  \,,\label{solHO}
\end{align}
is a solution for a Harmonic one. 

In particular, for the cases $m=1$ and $m=N+1$, we have
\begin{subequations}
\begin{align}
\Psi_{1}(Q,\Lambda)&= \prod_{c=1}^{N}\psi_c(Q_c,\Lambda) \,,\label{solm1}\\
\Psi_{N+1}(Q,\Lambda)&= \prod_{d=1}^{N}\chi_d(Q_d,\Lambda) \,, \label{solNplus1}
\end{align}
\end{subequations}
respectively. Notice that in these wave functions the parameters are $\Lambda= \{\Lambda^0,\Lambda^i\}$ \break $=\{t, X,Y,Z, B_1, \dots , B_N\}$ with $i=1,2,\dots, N+3$. Note also that the solution \eqref{eq:funciondeondaacoplados} can be written as \eqref{eq:funcionondagenacoplado}, by making the following identifications:
\begin{subequations}
    \begin{align}
    U^a_{m}(\Lambda)&= \left\{ \begin{array}{lcc}  U^{a}_{{}^\psi} \left(\Lambda\right) & \mathrm{if} &  a \geq m \,, \\ 
     U_{{}^\chi}^a(\Lambda) & \mathrm{if} & a< m \,,  \end{array} \right. \label{Uar}\\
    V^a_{m}(\Lambda)&= \left\{ \begin{array}{lcc} V_{{}^\psi}^a(\Lambda) & \mathrm{if} & a \geq m  \,, \\ V_{{}^\chi}^a(\Lambda) & \mathrm{if} & a< m \,,\end{array} \right. \\
    C_{m}(\Lambda)&=\prod_{a=m}^{N}C_{{}^\psi}^a(\Lambda) \prod_{a=1}^{m-1}C^a_{{}^\chi}(\Lambda)\,,
    \end{align}
\end{subequations}
where
\begin{subequations}
\begin{align}
    U_{{}^\psi}^a(\Lambda)&=\frac{B_a \omega_a \csc ^2\left( \omega_a t \right)}{B_a^2+\cot ^2\left( \omega_a t \right)}\,,\\
    V^a_{{}^\psi}(\Lambda)&=Y-\frac{\left(B_a^2-1\right) \omega_a \cot \left( \omega_a t \right)}{ B_a^2+\cot^2 \left( \omega_a t \right)}\, ,\\
    C_{{}^\psi}^a(\Lambda)&=\left( \frac{B_a \omega}{\pi} \right)^{1/4} \frac{1}{\sqrt{\cos \left( \omega_a t\right)+i B_a \sin \left( \omega_a t\right)}}\,, \\
    U_{{}^\chi}^a(\Lambda)&=\frac{ B_a \alpha_a \operatorname{csch}^2(\alpha_a  t)}{B_a^2+\coth ^2(\alpha_a  t)}\,,\\
    V_{{}^\chi}^a(\Lambda)&=Y-\frac{ \left(B_a^2+1\right) \alpha_a  \coth (\alpha_a  t)}{B_a^2+\coth ^2(\alpha_a  t)}\,,\\
    C^a_{{}^\chi}(\Lambda)&=\left( \frac{B_a \alpha}{\pi} \right)^{1/4} \frac{1}{\sqrt{\cosh \left( \alpha_a t\right)+i
   B_a \sinh \left( \alpha_a t\right)}}\,.
\end{align}
\end{subequations}

For the regions $m=1$, $m=2,\ldots, N$, and  $m=N+1$ the expectation value of the Hamiltonian is respectively given by 
\begin{subequations}
\begin{align}
\langle \hat{H}_{1} \rangle &= \sum_{d=1}^{N}\frac{(1+B_d^2)\omega_d}{4B_d}\,,\\
\langle \hat{H}_{m} \rangle &= \sum_{c=1}^{m-1}\frac{(-1+B_c^2)\alpha_c}{4B_c}+\sum_{d=m}^{N}\frac{(1+B_d^2)\omega_d}{4B_d}\,,\\
\langle \hat{H}_{N+1} \rangle &= \sum_{c=1}^{N}\frac{(-1+B_c^2)\alpha_c}{4B_c}\,.
\end{align}
\end{subequations}

\subsection{Particular case}

For the purpose of providing a specific example, let us  consider  $N=2$ and 
$$\boldsymbol{A}= \left(  \begin{array}{cc}
  2   & -1 \\
 -1    & 2
\end{array}
\right)\,,$$
which has eigenvalues $a_1=1$ and $a_2=3$. In this case, the Hamiltonian \eqref{hamiltonianoqpyfija} reduces to 
\begin{align}
    \hat{H}=&\frac{1}{2} \left(\hat{p}_1^2+\hat{p}_2^2+Y\left(\hat{q}_1 \hat{p}_1+\hat{p}_1 \hat{q}_1+\hat{q}_2 \hat{p}_2+\hat{p}_2 \hat{q}_2 \right) \right)+\frac{1}{2} \left(X+2Z\right)\left(\hat{q}_1^2+\hat{q}_2^2\right)-Z \hat{q}_1 \hat{q}_2\,.
    \label{particularcase}
\end{align}

As we have explained, depending on the value of the subset of the parameters  $\{ Y, X, Z \}$, we can consider three regions (once it is written in the normal modes): 
 \begin{itemize}
     \item Region 1: $\omega_1$ and $\omega_2$ are real.
     \item Region 2: $\alpha_1$ and $\omega_2$ are real.
     \item Region 3: $\alpha_1$ and $\alpha_2$ are real.
 \end{itemize}

Notice that there is no region such that $\omega_1$ and $\alpha_2$ are real. This is because, as mentioned previously, the squared normal frequencies are ordered increasingly. Accordingly, the Hamiltonians, in normal coordinates, for each of these regions are
\begin{subequations}
\begin{align}
    \hat{H}_{1}&=\frac{1}{2} \left( \hat{P}_1^2+\omega_1^2 \hat{Q}_1^2\right)+\frac{1}{2} \left( \hat{P}_2^2+\omega_2^2 \hat{Q}_2^2\right)\,, \\
    \hat{H}_{2}&=\frac{1}{2} \left( \hat{P}_1^2-\alpha_1^2 \hat{Q}_1^2\right)+\frac{1}{2} \left( \hat{P}_2^2+\omega_2^2 \hat{Q}_2^2\right)\,, \\
    \hat{H}_{3}&=\frac{1}{2} \left( \hat{P}_1^2-\alpha_1^2 \hat{Q}_1^2\right)+\frac{1}{2} \left( \hat{P}_2^2-\alpha_2^2 \hat{Q}_2^2\right)\,.
\end{align}    
\end{subequations}
Furthermore, using \eqref{eq:funciondeondaacoplados}, \eqref{solm1} and \eqref{solNplus1}, we get the solutions of the corresponding Schrödinger equations 
\begin{subequations} 
\begin{align}
\Psi_{1}(Q,\Lambda)&=\psi_1(Q_1,\Lambda)\psi_{2}(Q_2,\Lambda) \,,\\
\Psi_{2}(Q,\Lambda)&=\chi_1(Q_1,\Lambda)\psi_2(Q_2,\Lambda)\,,\\
\Psi_{3}(Q,\Lambda)&=\chi_1(Q_1,\Lambda)\chi_2(Q_2,\Lambda)\,,
\end{align}
\end{subequations}
and the expectation values
\begin{subequations}
    \begin{align}
        \langle \hat{H}_{1} \rangle &= \frac{(1+B_1^2)\omega_1}{4B_1} +\frac{(1+B_2^2)\omega_2}{4B_2}\,,\\
        \langle \hat{H}_{2} \rangle &= \frac{(-1+B_1^2)\alpha_1}{4B_1}+\frac{(1+B_2^2)\omega_2}{4B_2}\,,\\
        \langle \hat{H}_{3} \rangle &= \frac{(-1+B_1^2)\alpha_1}{4B_1}+\frac{(-1+B_2^2)\alpha_2}{4B_2}\,.
    \end{align}
\end{subequations}

Following the approach introduced in Section \ref{sec2}, we calculate the tQMT and tBerry curvature for the three regions, and taking $\{\Lambda^I\}=\{t, X, Y, Z, B_1, B_2\}$ with $I=0,1,\dots, 5$. We then examine whether there is any smoothness in these quantities as we move from one region to another. Some remarks for each region, in order, are as follows:
\begin{enumerate}
    \item In region 1, only trigonometric functions are present in the tQMT and tBerry curvature. In the case $B_1=B_2=1$, the components $g_{nl}$ with $(n,l=1,2,3)$ become time-independent. Also, the components $g_{0I}$ becomes zero at $B_1=B_2=1$
    \item In region 2, both trigonometric and hyperbolic functions appear in the tQMT and tBerry curvature.
    \item In region 3, only hyperbolic functions emerge in the tQMT and tBerry curvature.
\end{enumerate} 
Notably, the components $g_{00},g_{44}, g_{45}$, and $g_{55}$ are time-independent and  $g_{04},g_{05}$, and $F_{13}$ are zero, across all regions. Also, $g_{04}$ and $g_{05}$ vanish in agreement with the one degree of freedom case, where $g_{03}=0$. Generally, the resulting expressions for tQMT and tBerry curvature are too large to be explicitly written out; thus, we illustrate their behavior in the following figures.
\begin{figure}[H]
  \centering
  \begin{subfigure}{0.45\linewidth}{\includegraphics[width=0.9\linewidth]{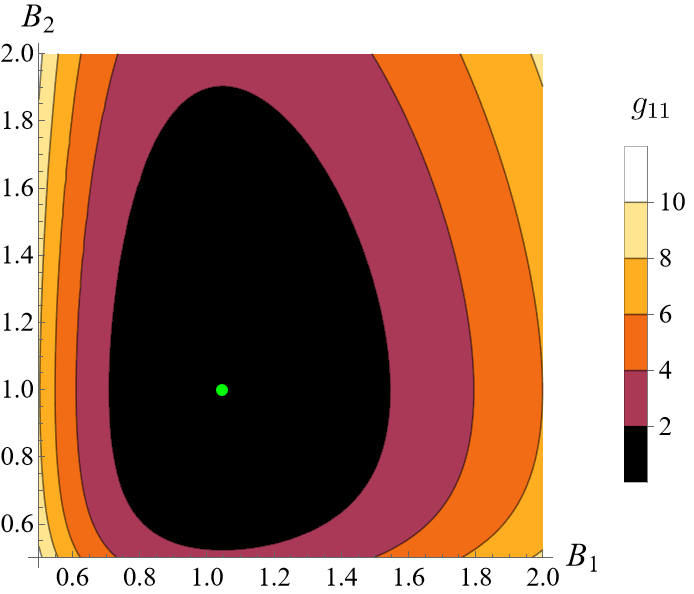}}
    \caption{Region 1}
    \label{fig:gabcopladovariandob1b2a}
  \end{subfigure}
  \begin{subfigure}{0.45\linewidth}\includegraphics[width=0.9\linewidth]{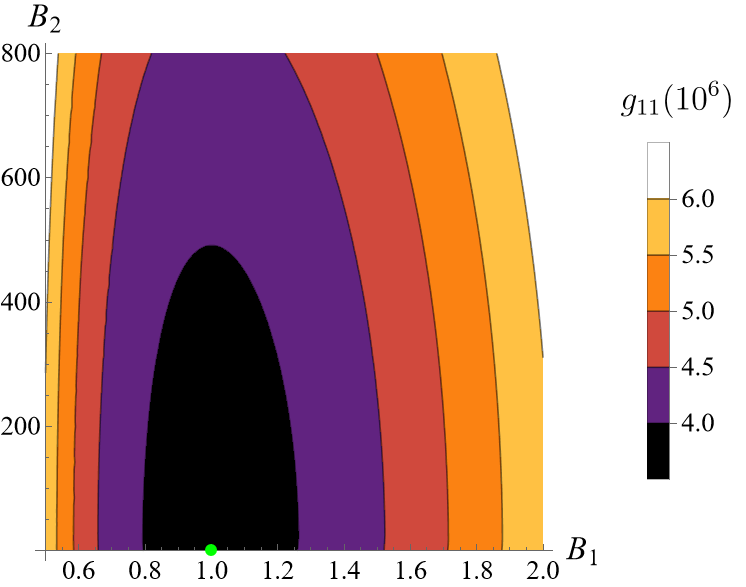}
    \caption{Region 2}
    \label{fig:gabcopladovariandob1b2b}  
  \end{subfigure}
   \begin{subfigure}{0.45\linewidth}\includegraphics[width=0.9\linewidth]{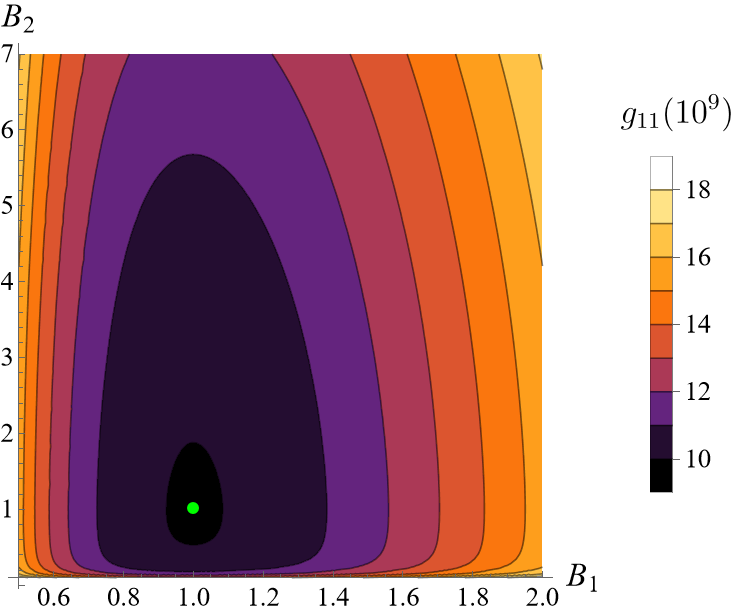}
    \caption{Region 3}
    \label{fig:gabcopladovariandob1b2c}  
  \end{subfigure}
  \captionsetup{font=small} 
  \caption{\justifying Component $g_{11}$ of the tQMT for the system \eqref{particularcase}. The parameters were fixed at $X=Z=1$ across all regions, while t was fixed as  (a) t=10, (b) t=5, (c) t=3.}    \label{fig:gabcopladovariandob1b2}
\end{figure}

\begin{figure}[H]
  \centering
  \begin{subfigure}{0.32\linewidth}{\includegraphics[width=0.9\linewidth]{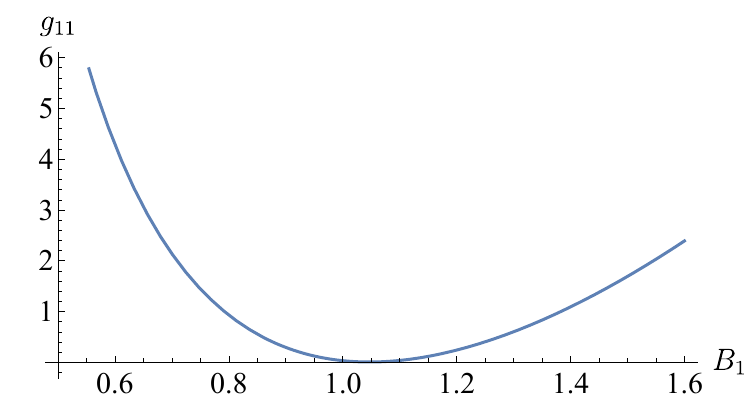}}
    \caption{Region 1}
    \label{fig:gabcopladovariandob2a}
  \end{subfigure}
  \begin{subfigure}{0.32\linewidth}\includegraphics[width=0.9\linewidth]{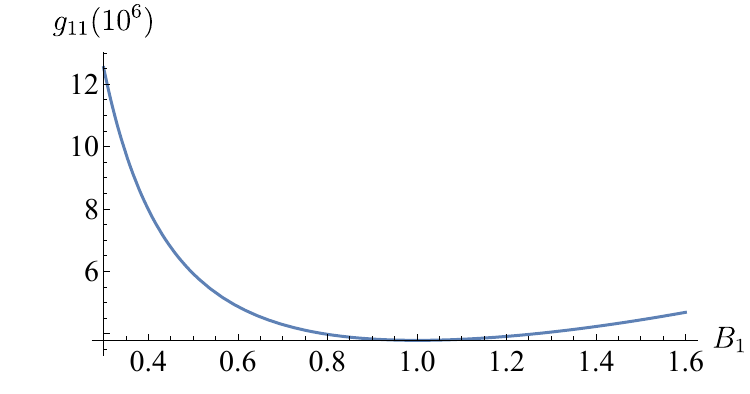}
    \caption{Region 2}
    \label{fig:gabcopladovariandob2b}  
  \end{subfigure}
   \begin{subfigure}{0.32\linewidth}\includegraphics[width=0.9\linewidth]{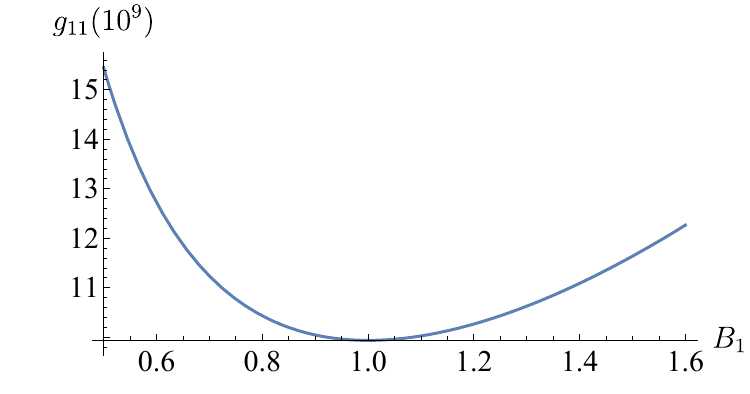}
    \caption{Region 3}
    \label{fig:gabcopladovariandob2c}  
  \end{subfigure}
  \captionsetup{font=small} 
  \caption{\justifying Component $g_{11}$ of the tQMT for the system \eqref{particularcase}. (a) Region 1, (b) Region 2, (c) Region 3.}    \label{fig:gabcopladovariandob1}
\end{figure}

In Fig. \ref{fig:gabcopladovariandob1b2}, we plot the component $g_{11}$ as a function of $B_1$ and $B_2$, while keeping the remaining parameters fixed at $X=Z=1$ across all regions.  With this choice of parameters, the value of $Y$ determines the region: region 1 corresponds to $Y\in(-\sqrt{2},\sqrt{2})$; region 2 to $Y\in(-2,-\sqrt{2})\cup (\sqrt{2},2)$; and region 3 to $Y\in(-\infty,-2)\cup(2,\infty)$. Taking into account this, the parameter $Y$ is selected as $Y=1,\sqrt{3},3$ for regions 1, 2, and 3, respectively. Furthermore, the parameter $t$ is chosen differently in each region due to the distinct growth ratios observed. Some comments regarding the Fig.~\ref{fig:gabcopladovariandob1b2} are:
\begin{itemize}
\item Region 1 displays the lowest growth ratio, while region 3 exhibits the highest growth ratio. Consequently, the choices for $t$ were set as $t=10,5,3$ for regions 1, 2, and 3, respectively. 
\item  In region 2, the scales for $B_1$ and $B_2$ differ because $B_1$ is associated with the inverted normal mode (exponential behavior) while $B_2$ with a harmonic normal mode (trigonometric behavior).

\item The three regions have a global minimum, which occurs in the neighborhood of $B_1=B_2=1$ and is represented by a green point. This can be better seen in Fig.~\ref{fig:gabcopladovariandob1}, where we plot the metric component $g_{11}$ as a function of $B_1$ while keeping $B_2$ fixed at 1.
\end{itemize}

Similar behavior is observed for the metric components involving subscripts $1,2,3$ (i.e., those obtained by varying the parameters $X, Y, Z$).

\subsubsection{Scalar curvature}

For a more comprehensive analysis of the tQMT, we compute the scalar curvatures $\mathcal{R}[\Lambda^{I},\Lambda^{J}]$ and $\mathcal{R}[\Lambda^{I},\Lambda^{J},\Lambda^{K}]$ associated with the $2\times 2$ metrics $\tilde{g}[\Lambda^I,\Lambda^J]$ and $3\times 3$ metrics  $\tilde{g}[\Lambda^I,\Lambda^J,\Lambda^K],$ respectively. Once again, the expressions are too lengthy to be explicitly written out. Therefore, we illustrate their behavior graphically. The parameter values remain consistent with those used for the plots of components of the metric (see Fig.~\ref{fig:gabcopladovariandob1b2}), except for particular choices involving $t$, which will be explicitly indicated in each figure.

\paragraph{Scalar curvature taking the parameters $X$ and $B_1$ \newline}

\begin{figure}[H]
  \centering
  \begin{subfigure}{0.45\linewidth}{\includegraphics[width=0.9\linewidth]{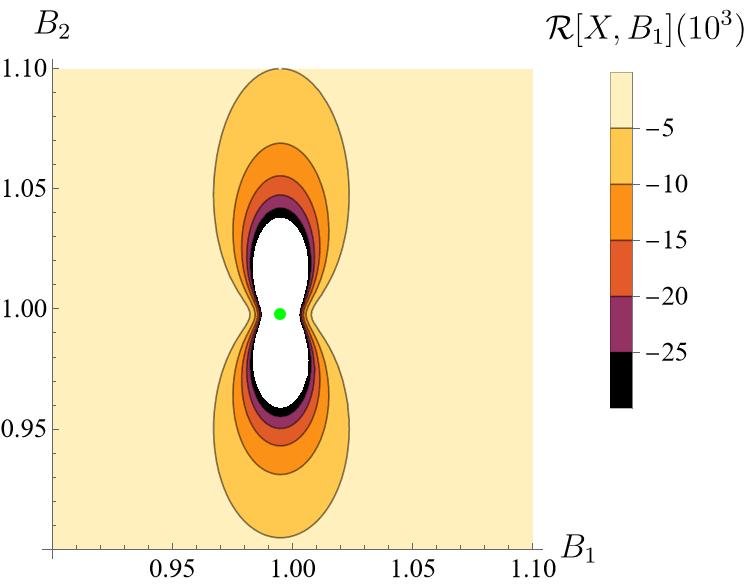}}
    \caption{Region 1}
    \label{fig:R14b1b2a}
  \end{subfigure}
  \begin{subfigure}{0.45\linewidth}\includegraphics[width=0.9\linewidth]{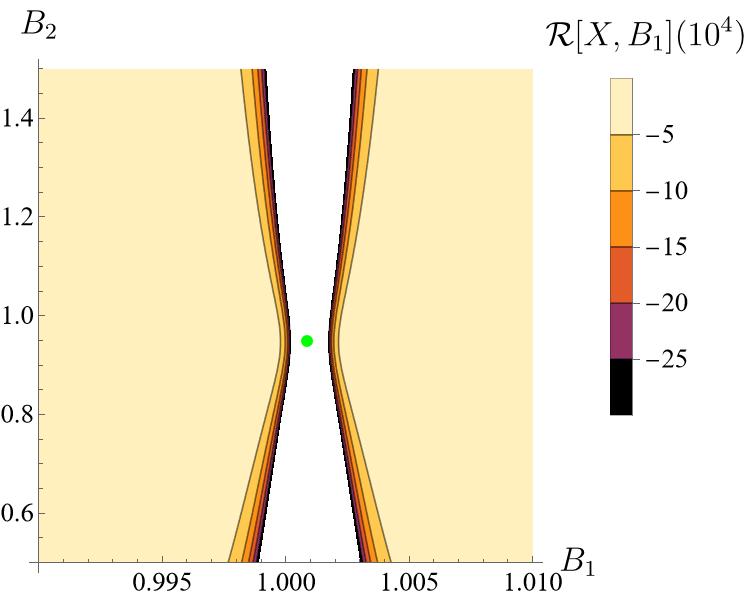}
    \caption{Region 2}
    \label{fig:R14b1b2b}  
  \end{subfigure}
   \begin{subfigure}{0.45\linewidth}\includegraphics[width=0.9\linewidth]{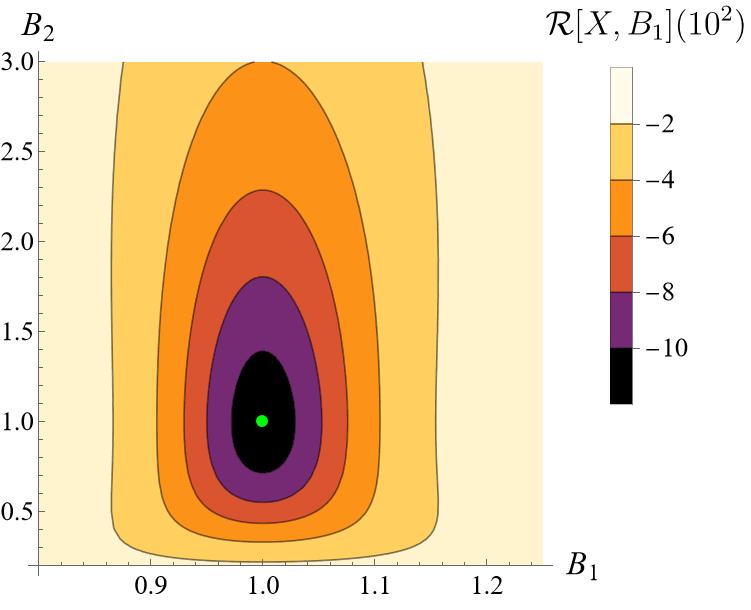}
    \caption{Region 3}
    \label{fig:R14b1b2c}  
  \end{subfigure}
  \captionsetup{font=small} 
  \caption{\justifying Scalar curvature $\mathcal{R}[X, B_1]$  as a function of the parameters $B_1$ and $B_2$. As in Fig. \ref{fig:gabcopladovariandob1b2}, the parameters were fixed at $X=Z=1$ across all regions, while t was fixed as  (a) t=10, (b) t=5, (c) t=3. }    \label{fig:R14b1b2}
\end{figure}

In Fig. \ref{fig:R14b1b2}, we show the scalar curvature $\mathcal{R}[X, B_1]$  as a function of the parameters $B_1$ and $B_2$. To provide clarity, values of $\mathcal{R}[X, B_1]$ smaller than $-30\times 10^3$ and $-30\times 10^4$ are represented by white in sub-figures (a) and (b) respectively. Once again, the minimum (green point) is observed to be near $B_1=B_2=1$. Also, the neighborhood of the minimum is the region where the scalar curvature exhibits the most pronounced variation. Therefore, in this neighborhood, a small change in the parameters $B_1$ and $B_2$ produces the greatest variation in the quantum system.

\begin{figure}[H]
  \centering
    \begin{subfigure}{0.45\linewidth}\includegraphics[width=0.9\linewidth]{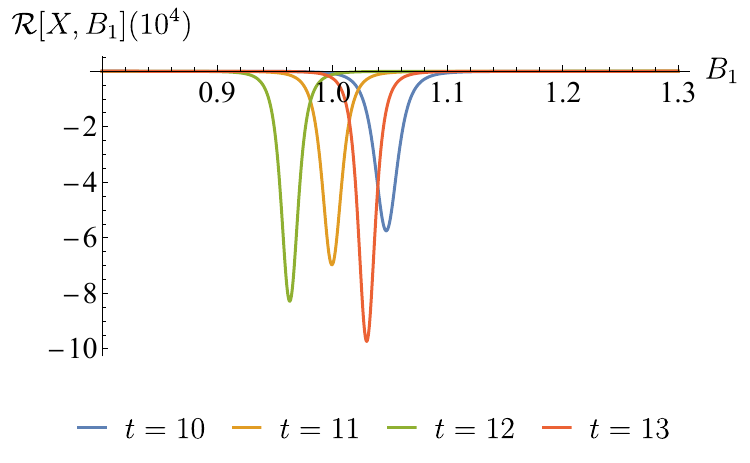}
    \caption{}
    \label{fig:r14ta}
  \end{subfigure}
  \begin{subfigure}{0.45\linewidth}\includegraphics[width=0.9\linewidth]{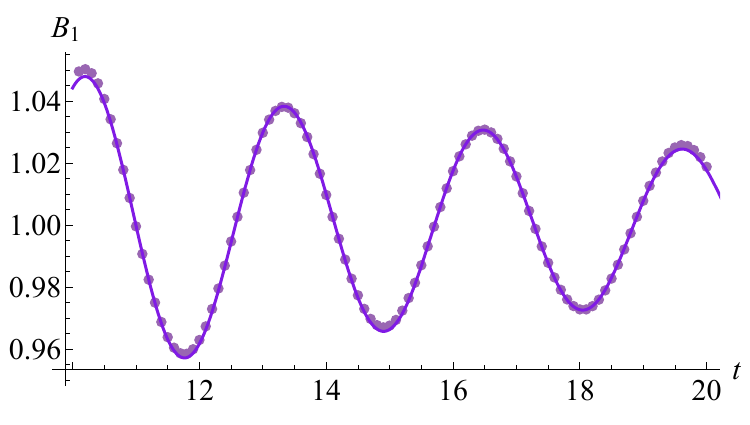}
  \caption{}
  \label{fig:r14tb} 
  \end{subfigure}
  \captionsetup{font=small} 
  \caption{\justifying a)  Scalar curvature $\mathcal{R}[X,B_1]$, it has an oscillating minimum around $B_1=1$. b) Minima points of the scalar curvature in region 1 as a function of $B_1$ while fixing $B_2$. }  
   \label{fig:r14t}
\end{figure}

{\bf {Region 1}}. To observe the variation of the minimum of the scalar curvature over time, in Region 1, we cut Fig. \ref{fig:R14b1b2a} at $B_2=1$ and evaluate for $t=10,11,12,13$. The resulting plot is presented in Fig. \ref{fig:r14ta}, where it can be observed that the minima of $\mathcal{R}[X, B_1]$  shift as $t$ varies.

In Fig.~\ref{fig:r14ta}, we can see that the minimum seems to oscillate around $B_1=1$. This oscillatory behavior can be explicitly appreciated in Fig.~\ref{fig:r14tb}, where we plot the value of $B_1$ at which the minimum of  $\mathcal{R}[X, B_1]$ occurs, as a function of $t$, with steps of $\Delta=0.1$. Plotting these points as a function of time reveals a damped oscillator behavior with a constant period. The function used to fit the data is
\begin{align}
    f(t)=c_0\sin{(c_1 +c_2 t)}\exp(-c_3t)+1\,,
    \label{seno}
\end{align}
obtaining $c_0= -0.098958$, $c_1=  0.0204918$, $c_2 = -2.001312$, $c_3 = 0.071122$, with a coefficient of determination $R^2=0.999999$. Note that the limit $t\to\infty$ of \eqref{seno} is 1, and  hence the minimum of $\mathcal{R}[X,B_1]$ in the limit $t\to \infty$ occurs at $B_1=B_2=1$. 

\begin{figure}[H]
  \centering
    \begin{subfigure}{0.45\linewidth}{\includegraphics[width=0.9\linewidth]{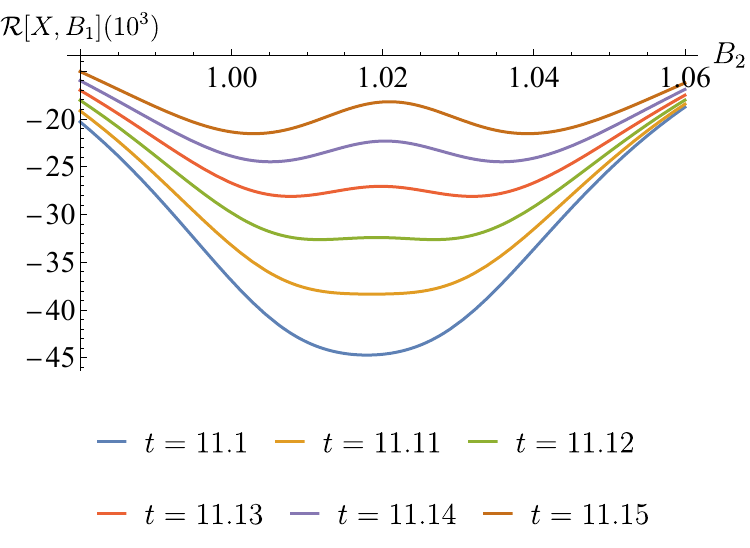}}
    \caption{}
    \label{fig:r14b2a}
  \end{subfigure}
  \begin{subfigure}{0.45\linewidth}\includegraphics[width=0.9\linewidth]{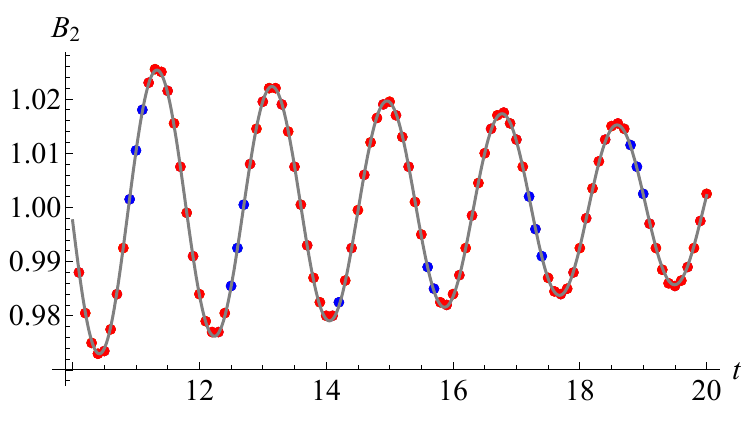}
    \caption{}
    \label{fig:r14b2b}
    \label{fig:minimosr1b}  
  \end{subfigure}
  \captionsetup{font=small} 
  \caption{\justifying a) Scalar curvature $\mathcal{R}[X, B_1]$ for different times. The critical points change from minimums to local maximums.  b) Extremal points of $\mathcal{R}[X, B_1]$ in region 1 as a function of $B_2$ while fixing $B_1$. }  
    \label{fig:r14b2}
\end{figure}

A similar analysis can be carried out for fixed  $B_1$ while varying $B_2$. In Fig. \ref{fig:r14b2a}, we plot $\mathcal{R}[X, B_1]$ as a function of $B_2$, for different values of $t$. Here we can see that $\mathcal{R}[X, B_1]$ exhibits a minimum for  $t=11.11, 11.12$, while it has a local maximum surrounded by two minimums for $t=11.13,11.14,11.15$. Then, we can identify two different patterns. The first one consists of minima in the region close to $B_2=1$ and the second involves a local maximum near $B_2=1$, surrounded by a minimum on its left and another on the right. In Fig.~\ref{fig:r14b2b}, we plot the value of $B_2$ at which the minimum or local maximum of  $\mathcal{R}[X, B_1]$ occurs as a function of $t$, with steps of $\Delta=0.1$. Here, the red points correspond to maxima of $\mathcal{R}[X, B_1]$, while the blue ones correspond to minima. The function used to fit the data is \eqref{seno}, with $c_0 =-0.056185$, $c_1 = -3.124388$, $c_2 =3.462878$, $c_3 =  0.070285$.

\begin{figure}[H]
  \centering
  \begin{subfigure}{0.45\linewidth}{\includegraphics[width=0.9\linewidth]{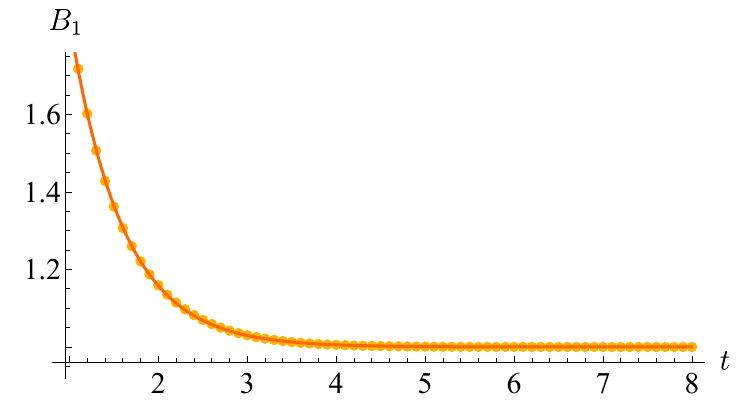}}
    \caption{}
    \label{fig:minimosr2a}
  \end{subfigure}
  \begin{subfigure}{0.45\linewidth}\includegraphics[width=0.9\linewidth]{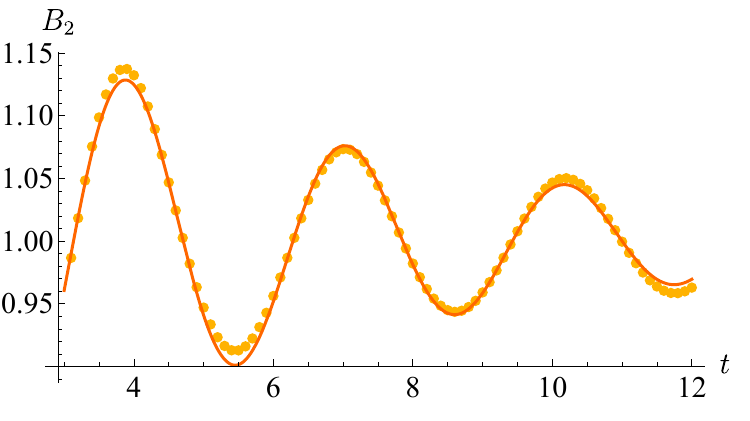}
    \caption{}
    \label{fig:minimosr2b}  
  \end{subfigure}
  \captionsetup{font=small} 
  \caption{\justifying Minima points of the scalar curvature in region 2 as a function of: (a)  $B_1$ while fixing $B_2$, (b) $B_2$ while fixing $B_1$. }  
    \label{fig:minimosr2}
\end{figure}

{\bf {Region 2}}. We examine the minima of $\mathcal{R}[X, B_1]$ when cutting Fig. \ref{fig:R14b1b2b} at $B_1=1$ for different times.  Additionally, we study the alternative scenario where we cut Fig. \ref{fig:R14b1b2b} at  $B_2=1$. We consider both scenarios due to the distinct behavior of the normal modes: the normal mode associated with $B_1$ behaves as an inverted oscillator, whereas the one associated with $B_2$ behaves as a harmonic oscillator. In Fig.~\ref{fig:minimosr2a}, we plot the value of $B_1$ at which the minimum of  $\mathcal{R}[X,B_1]$ occurs, as a function of $t$ and taking $B_2=1$.The function used to fix the data in this case is
\begin{align}
    f(t)=c_0\sinh{(c_1 +c_2 t)}\exp(-c_3t)+1\,.
    \label{senohip}
\end{align}
In Fig.~\ref{fig:minimosr2b}, we now plot the value of $B_2$ at which the minimum of  $\mathcal{R}[X, B_1]$ occurs, as a function of $t$ and considering $B_1=1$. In this scenario, we fit the data using \eqref{seno}.

\begin{figure}[H]
  \centering
  \begin{subfigure}{0.45\linewidth}{\includegraphics[width=0.9\linewidth]{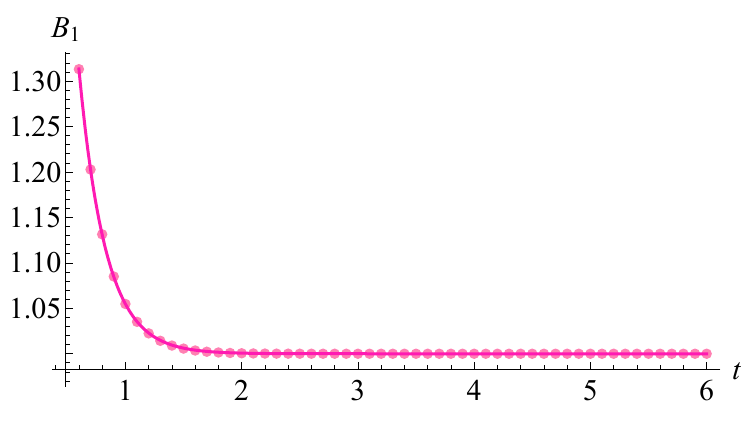}}
    \caption{}
    \label{fig:minimosr3a}
  \end{subfigure}
  \begin{subfigure}{0.45\linewidth}\includegraphics[width=0.9\linewidth]{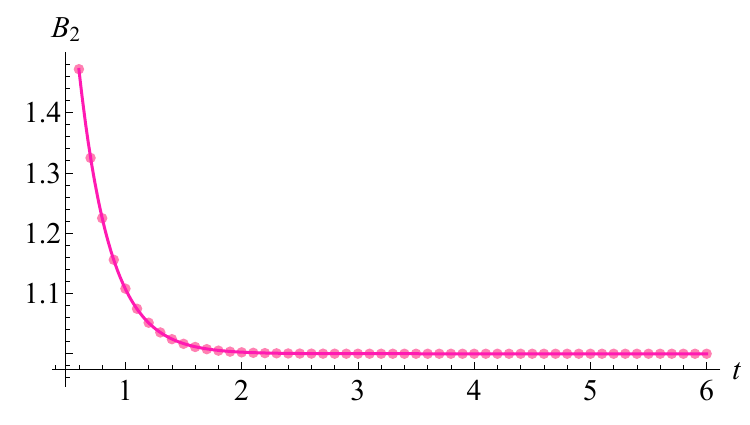}
    \caption{}
    \label{fig:minimosr3b}  
  \end{subfigure}
  \captionsetup{font=small} 
  \caption{\justifying Extremal points of the scalar curvature $\mathcal{R}[X,B_1]$ in region 3 as a function of (a)  $B_1$ while fixing $B_2$ and (b) $B_2$ while fixing $B_1$. }  
    \label{fig:minimosr3}
\end{figure}

{\bf {Region 3}}. We analyze the minima of $\mathcal{R}[X,B_1]$ by first setting $B_1=1$ and then $B_2=1$ in Fig. \ref{fig:R14b1b2c}, for different values of $t$. In Figs. \ref{fig:minimosr3a} and \ref{fig:minimosr3b}, we plot the value of $B_1$ and $B_2$, respectively, at which the minimum of $\mathcal{R}[X,B_1]$ occurs, as a function of $t$. In these scenarios, the normal modes associated with $B_1$ and $B_2$ behave as inverted oscillators, and consequently, both sets of data can be fixed using \eqref{senohip}, as we show in Fig. \ref{fig:minimosr3}.

\paragraph{Scalar curvature considering the rest of the parameters \newline}

Although the analysis was explicitly done for $\mathcal{R}[X, B_1]$, a similar examination can be carried out for the other scalar curvatures $\mathcal{R}[\Lambda^I,\Lambda^J]$. Before beginning the analysis, we introduce two new notations. First, we denote the extremal points of $\mathcal{R}[\Lambda^I,\Lambda^J]$ as $x^*_{IJ}$. Second, we introduce the normalized scalar curvature over $\Omega$,  defined as 
\begin{align}
R_{\text{norm}}[\Lambda^I,\Lambda^J]=\frac{\mathcal{R}[\Lambda^I,\Lambda^J]}{\max_{\Omega}\mathcal{R}[\Lambda^I,\Lambda^J]-\min_{\Omega}\mathcal{R}[\Lambda^I,\Lambda^J]} \,,
\end{align}
where $\Omega$ is a sub-manifold embedded in the time-parameter manifold $\mathcal{M}$ and $\max_{\Omega}\mathcal{R}[\Lambda^I,\Lambda^J]$ ($\min_{\Omega}\mathcal{R}[\Lambda^I,\Lambda^J]$) is the maximum (minimum) of $\mathcal{R}[\Lambda^I,\Lambda^J]$ at the region $\Omega$.  We introduce $R_{\text{norm}}$ due to the significant variation in the scales of $\mathcal{R}$ for different parameter choices. Normalizing the scalar curvature allows a consistent comparison across different parameter choices, mitigating the effects of scale variation. Since in our plots, we fix all the parameters, excepting $B_2$, the sub-manifold $\Omega$ reduces to an open interval which in our example is $(0.9,1.05)$. There are some important remarks for the scalar curvatures, which hold for all regions:

\begin{itemize}
    \item The scalar curvatures $\mathcal{R}[X,Z]$ and $\mathcal{R}[B_1,B_2]$ are zero.
    \item The scalar curvatures coming from the variation of parameters $t,B_1,B_2$ are time-independent.
\end{itemize}

\begin{figure}[H]
  \centering
  \begin{subfigure}{0.32\linewidth}{\includegraphics[width=0.9\linewidth]{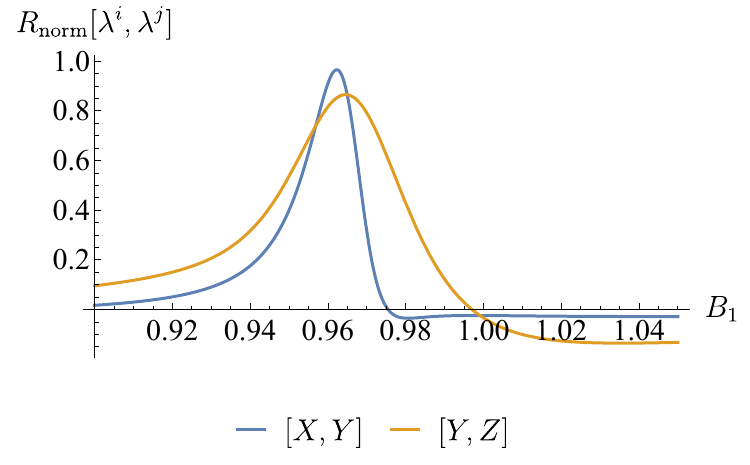}}
    \caption{}
    \label{fig:Rijregion1}
  \end{subfigure}
  \begin{subfigure}{0.32\linewidth}\includegraphics[width=0.9\linewidth]{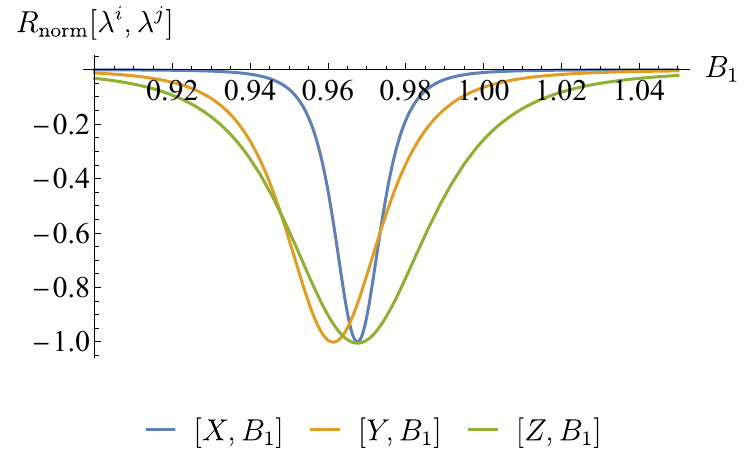}
    \caption{}
    \label{fig:Rib1region1}  
  \end{subfigure}
  \begin{subfigure}{0.32\linewidth}\includegraphics[width=0.9\linewidth]{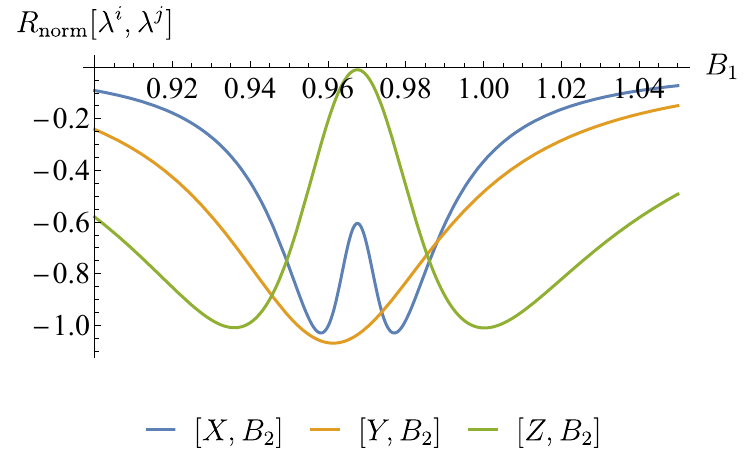}
    \caption{}
    \label{fig:Rib2region1}  
  \end{subfigure}  
  \captionsetup{font=small} 
  \caption{\justifying Normalized scalar curvatures $R_{\text{norm}}[\lambda^i,\lambda^j]$ in Region 1. They present an extremal point in the vicinity of $B_1=1$. The parameters were fixed at $X=Y=Z=1$ and $t=15$. }   
    \label{fig:Rregion1}
\end{figure}

The scalar curvatures $\mathcal{R}[\lambda^i,\lambda^j]$ (excepting $\mathcal{R}[X, Z]$ and $\mathcal{R}[B_1, B_2]$), exhibit behavior analogous to that of $\mathcal{R}[X, B_1]$, i.e, they have an extremal point in the vicinity of $B_1=1, B_2=1$. Even the extremal points appear to be very close to each other, i.e., $x^*_{i_1j_1}\approx x^*_{i_2j_2}$ for equal choices of parameters.  This can be seen in Fig. \ref{fig:Rregion1} for region 1. Similar behavior occurs in regions 2 and 3. 

Now, let us analyze the remaining scalar curvatures, which are of the form $\mathcal{R}[t,\lambda^i]$. In this case, we need to distinguish between regions 1, 2, 3.

\begin{figure}[h!]
    \centering
    \includegraphics[width=0.55\linewidth]{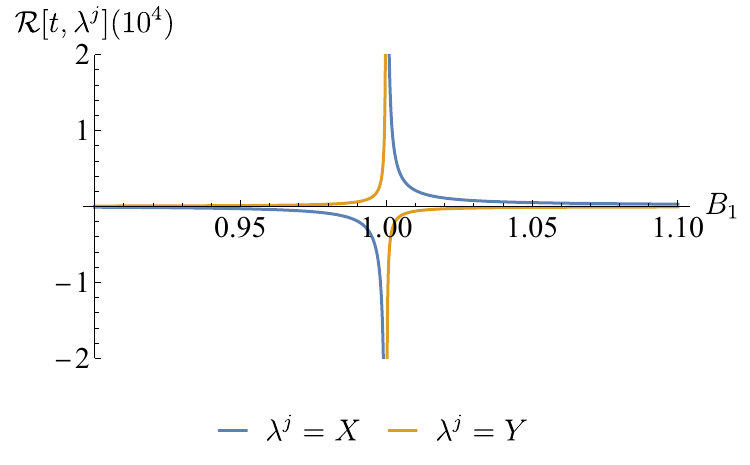}
    \captionsetup{font=small} 
  \caption{\justifying Scalar curvatures $R[t,X]$ and $R[t,Y]$ in region 1. They show a singularity at $B_1=B_2=1$. The parameters were fixed at $B_2=X=Y=Z=1,$ and $t=15$. A similar behavior is observed by fixing $B_1$ and varying $B_2$.}  \label{fig:Rtjregion1}
\end{figure}

Region 1. $\mathcal{R}[t,\lambda^i]$ is singular at $B_1=B_2=1$, which is a consequence of the fact that the determinant of the metric $\tilde{\boldsymbol{g}}[t,\lambda^i]$ is zero under this choice of parameters. This behavior can be observed in Fig. \ref{fig:Rtjregion1}, where we plot $R[t,X]$ and $R[t,Y]$ as functions of $B_1$. On the other hand, the scalar curvatures $\mathcal{R}[t, B_1]$ and $\mathcal{R}[t, B_2]$ result in non-continuous functions at $B_1=B_2=1$, whose value depends on the order in which the limit is taken. 

\begin{figure}[h!]
    \centering
    \includegraphics[width=0.55\linewidth]{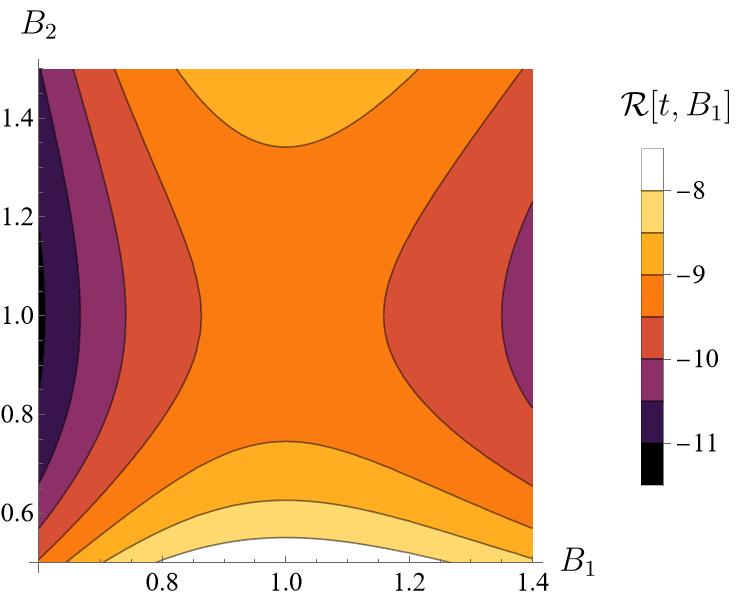}
     \captionsetup{font=small} 
  \caption{\justifying Scalar curvature $\mathcal{R}[t, B_1]$ in region 3. The parameters were fixed at $X=Y=Z=1$ and $t=3$. It has a saddle point at $B_1=B_2=1$. In region 2, this curvature shows an analogous behavior. }    \label{fig:R04region3}
\end{figure}

Regions 2 and 3. The scalar curvatures $\mathcal{R}[t,B_1]$ and $\mathcal{R}[t,B_2]$ exhibit behavior similar to that of $\mathcal{R}[\lambda^i,\lambda^j]$. However, for these scalar curvatures, their extremal point occurs precisely at $B_1=B_2=1$. The extremal points are saddle points, as illustrated in Fig. \ref{fig:R04region3}.

After analyzing the scalar curvatures of the previous metrics, we conclude that for sufficiently large $t$, there is an extremal point in the neighborhood of $B_1=B_2=1$. The corresponding scalar curvatures undergo the most abrupt changes around this point. Consequently, variations in the parameters $B_1$ and $B_2$ in this vicinity result in the most significant changes in the quantum system.

\subsection{Purity}

For systems with $N\geq 2$ degrees of freedom, we can compute additional non-trivial quantities of physical interest, especially those related to quantum entanglement. For instance, the purity $\mu$, which is solely determined by the covariance matrix for Gaussian states \cite{Agarwal1971, Holevo1999, Dodonov_2002, Paris2003, deGosson2006, diosi2011, Golubeva2014, serafini2017, deGosson2019, Demarie2018}. Specifically, for a given quantum state  $\ket{\Psi(\Lambda)}$, the quantum covariance matrix $\boldsymbol{\sigma} = (\sigma_{\alpha \beta})$ is a $2N\times 2N$ matrix, with components given by
\begin{align}
\sigma_{\alpha \beta} = \frac{1}{2} \langle \hat{r}_{\alpha} \hat{r}_{\beta} + \hat{r}_{\beta} \hat{r}_{\alpha}  \rangle-\langle \hat{r}_{\alpha}\rangle \langle \hat{r}_{\beta} \rangle\,, \label{qumet}
\end{align}
where $\hat{r}=\{\hat{r}_{\alpha}\}=(\hat{q}_1,\dots,\hat{q}_N,\hat{p}_1,\dots,\hat{p}_N)^{\intercal}$ ($\alpha,\, \beta =1,\dots,2N$) is a $2N$-dimensional column vector  and $ \langle \hat{O} \rangle=\langle  \Psi(\Lambda) | \hat{O} | \Psi(\Lambda) \rangle$ stands for the expectation value of an operator $\hat{O}$.
Then, considering an $k$-mode Gaussian state ($k$ denotes the number of degrees of freedom of the subsystem formed by the particles $a_1,\dots,a_k$ of the system of $N$ degrees of freedom) with reduced quantum covariance matrix $\boldsymbol{\sigma}^{(k)}$, the purity is
\begin{align}
\mu \left(a_1,a_2,\dots,a_k\right)&= \frac{1}{2^k} \frac{1}{\sqrt{\det \boldsymbol{\sigma}^{(k)} }}\,. \label{eq:qpu}
\end{align}

As we remark, the states \eqref{eq:funcionondagenacoplado}  we consider in the present article are Gaussian, even if they have energy dispersion. To compute the purity of these states, we begin by calculating the corresponding covariance matrix. It is convenient to express the original position and momentum operators in terms of the operators corresponding to the normal modes
\begin{subequations}\label{covarianzagen}
\begin{align}
\hat{q}_a& =\sum_b S_{ab} \hat{Q}_b\,,\\
\hat{p}_a&=\sum_b \left(S_{ab}\hat{P}_b-Y S_{ab}\hat{Q}_b\right)\,.
\end{align}
\end{subequations}

Considering a wave function of the form \eqref{eq:funcionondagenacoplado}, the averages involved in the covariance matrix are
\begin{subequations}\label{promedios}
\begin{align}
\langle \hat{q}_a \hat{q}_b \rangle&= \tfrac{1}{2}\left(\boldsymbol{S} \boldsymbol{U}^{-1}_m\boldsymbol{S}^{\intercal} \right)_{ab}\,,\\
\frac{1}{2}\langle \hat{q}_a \hat{p}_b+\hat{p}_b \hat{q}_a \rangle&= \tfrac{1}{2}\left(Y \boldsymbol{S} \boldsymbol{U}^{-1}_{m}\boldsymbol{S}^{\intercal}\right)_{ab}\,,\\
\langle \hat{p}_a \hat{p}_b \rangle&= \tfrac{1}{2}\left(\boldsymbol{S}\boldsymbol{U}_m\boldsymbol{ S}^{\intercal} + Y^2\boldsymbol{S} \boldsymbol{U}^{-1}_m\boldsymbol{S}^{\intercal} \right)_{ab}\,, \label{18c}\\
\langle \hat{q}_a \rangle&= \langle \hat{p}_a \rangle=0\,,
\end{align}
\end{subequations}
where $ \boldsymbol{U}_m=\operatorname{diag} U^a_m$ with $U^a_m$ being defined in \eqref{Uar}. 

In general, to calculate the purity of a subsystem consisting of the first $k$ oscillators, immersed in the chain of $N$ oscillators (with $k \leq N$), we need to obtain the reduced covariance matrix in region $m$. This quantity is denoted by $\boldsymbol{ \sigma}^{(k)}_{m}$, and for the present example it is given by
\begin{align}\label{redcovex2}
\boldsymbol{ \sigma}^{(k)}_{m}=\left(\begin{array}{cc} \boldsymbol{D}_m & -Y\boldsymbol{D}_m  \\- Y\boldsymbol{D}_m & \boldsymbol{L}_m+Y^2 \boldsymbol{D}_m\end{array}\right)\,,
\end{align}
where $\boldsymbol{L}_m$ and $\boldsymbol{D}_m$ are $k\times k$ submatrices of 

\begin{subequations}
\begin{align}
\boldsymbol{S} \boldsymbol{U}_m\boldsymbol{S}^{\intercal}&=\left(\begin{array}{cc} \boldsymbol{L}_m & \boldsymbol{B}_m \\ \boldsymbol{B}^{\intercal}_m & \boldsymbol{C}_m\end{array}\right)\,,\\
\boldsymbol{S} \boldsymbol{U}_m^{-1}\boldsymbol{S}^{\intercal}&=\left(\begin{array}{cc} \boldsymbol{D}_m & \boldsymbol{E}_m \\ \boldsymbol{E}^{\intercal}_m & \boldsymbol{F}_m\end{array}\right)\,,
\end{align}
\end{subequations}
respectively.

Using \eqref{redcovex2}, the  purity \eqref{eq:qpu} of the $k$ oscillators in region $m$ turns out to be 
\begin{align}\label{classpurity}
\mu_m (a_1,\dots,a_k)=\dfrac{1}{\sqrt{\det \boldsymbol{L}_m \det \boldsymbol{D}_m }}\, .
\end{align}
We must emphasize that this purity is in general time-dependent.

In particular, for the system of two oscillators, the purity of the reduced system of one oscillator in the region $m$ is given by
\begin{align}
\mu_m(a_1)=\frac{2\sqrt{U_{m}^1U^2_{m}}}{U^1_{m}+U^2_{m}}\,,
\label{Puritydos}
\end{align}
where ${m}={ 1,2,3}$. The expressions for the purity $\mu_m(a_1)$, for each region $m$, simplifies at $B_1=B_2=1$ and reduce to
\begin{subequations}
\begin{align}
\mu_1(a_1) &=\frac{2\sqrt{\omega_1\omega_2}}{\omega_1+\omega_2}\,,\\
\mu_2(a_1) &=\frac{2\sqrt{\sec(2 \alpha_1 t)\alpha_1\omega_2}}{\sec(2 \alpha_1 t)\alpha_1+\omega_2}\,,\\
\mu_3(a_1) &=\frac{2\sqrt{\sec(2 \alpha_1 t)\sec(2\alpha_2 t)\alpha_1\alpha_2}}{\sec(2 \alpha_1 t)\alpha_1+\sec(2 \alpha_2 t)\alpha_2} \,.
\end{align}
\end{subequations}
Notice that for this election, $B_1=B_2=1$, in region 1, the state is time-independent, and then, the purity is also time-independent. This purity coincides with the result obtained in \cite{Diaz2022}.

Let us now analyze the behavior of the purity at the two bifurcations. The first bifurcation appears at the point $X - Y^2 + Z a_1 = 0$, and it is the one that divides regions 1 and 2; while the second bifurcation occurs at $X - Y^2 + Z a_2 = 0$, and it is the one that divides regions 2 and 3. Therefore, to analyze them, it is necessary to calculate the limit from the left and the right of the resulting purities. For region 1, this is equivalent to taking the limit $\omega_1 \to 0$ for $\mu_1(a_1)$, and $\alpha_1 \to 0$ for $\mu_2(a_1)$. Analogously, for region 2, this is equivalent to taking the limit $\omega_2 \to 0$ for $\mu_2(a_1)$, and $\alpha_2 \to 0$ for $\mu_3(a_1)$. To achieve this, we calculate $U_{{}^\psi}^a$ and $U_{{}^\chi}^a$ in the limit $\omega_a\to 0$ and $\alpha_a\to 0$, respectively, obtaining
\begin{subequations}
\begin{align}
    \lim_{\omega_a\to 0}U_{{}^\psi}^a\to 0 \,,\\
    \lim_{\alpha_a\to 0} U_{{}^\chi}^a\to 0\,.
\end{align}
\end{subequations}
Since we are considering different frequencies, we can infer from equation \eqref{Puritydos} that the numerator approaches zero, while the denominator remains finite. Hence, the purity at the bifurcation points is zero, meaning the system is completely entangled. Also, since the left and right limits coincide, we conclude that the purity is continuous, however, a straightforward calculation shows that it is not differentiable. 

\begin{figure*}
\centering
\includegraphics[scale=.45]{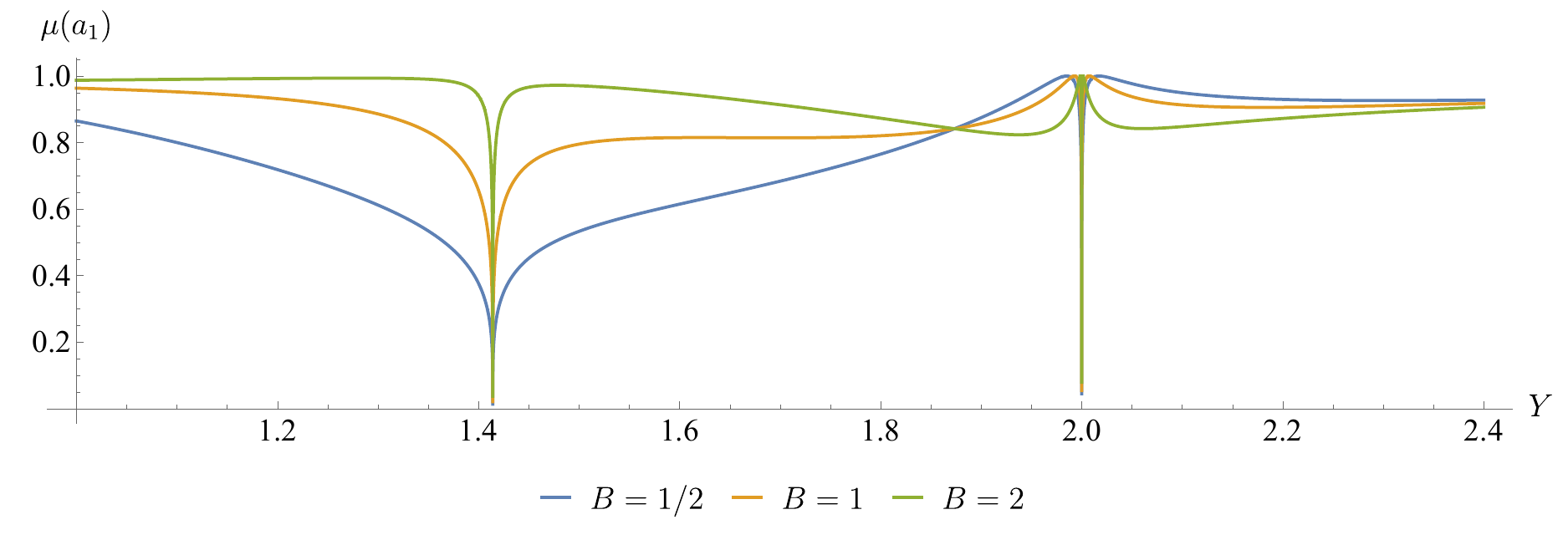}
  \caption{\justifying Purity as a function of the parameter $Y$. The parameters are fixed at $X=Z=1$, with this choice of parameters, the bifurcation points are $Y=\sqrt{2}$ and $Y=2$.}
    \label{fig:purity}
\end{figure*}

For the model introduced in \eqref{particularcase}, the previous analysis can be observed in Fig. \ref{fig:purity}, where we fix the parameters $X=Z=1$, and we consider $B_1=B_2=B$.  With this choice of parameters, the bifurcation points are $Y=\{\sqrt{2},2\}$.

Let us refer back to equation \eqref{Puritydos} in region 1. Remembering that our time-dependent Gaussian functions \eqref{eq:funcionondagenacoplado} can be expressed as a Fourier series expansion of the even eigenfunctions of the harmonic oscillator as was proven \eqref{27}, we can rewrite the purity as follows
\begin{align}
\label{generadora}
   \mu_1(a_1)=\!\!\!\!\!\sum_{\{ 2m,2n\} } \!\!\!\!\!C_{\{2m,2n\}}\!\!\! \int \!\! dq_1 dq_2 dx_1 dx_2 f_{2m2n}(x_1,x_2,q_1,q_2) \,,
\end{align}
with $C_{\{2m,2n\}}>0,$ $m=(m_1,m_2, m_3,m_4)$,  $n=(n_1,n_2, n_3,n_4)$, and
\begin{align}
     f_{mn}(x_1,x_2,q_1,q_2)= u_{n_1,m_1}(x_1,q_2)u_{n_2,m_2}(x_1,x_2)u_{n_3,m_3}(q_1,q_2)u_{n_4,m_4}(q_1,x_2)\,,
\end{align}
where $u_{m_1,m_1}$ is the wave function of the eigenstate with energy $E_{n_1,m_1}=\omega_1( m_1+1/2)+\omega_2(n_2+1/2)$. In particular, if $m_1=m_2=m_3=m_4$ and $n_1=n_2=n_3=n_4$, the purity of the excited eigenstate with energy $E_{n_1,m_1}$ is given by
\begin{align}
\label{purezaexcitados}
\mu_{m_1n_1} (a_1) : =   \int dq_1 dq_2 dx_1 dx_2 f_{m_1 n_1}(x_1,x_2,q_1,q_2)\,.
\end{align}
In other words, equation \eqref{Puritydos} encapsulates all the information about the purity of all even states. Notably, we show that $\mu_1(a_1)\to 0$ at the bifurcation point. From equation \eqref{purezaexcitados}, we can conclude that the purity of the reduced system of one oscillator embedded in a system of two oscillators of any even excited eigenstate at the bifurcation points is zero.

\subsubsection{Equal frequencies}

Now, let us analyze the case where the frequencies are equal. In this case, the parameter space is divided only into 2 regions instead of 3. In the first one, both modes exhibit harmonic behavior, while in the second both modes exhibit inverted oscillator behavior.

The expression for purity remains as \eqref{Puritydos} for both regions. On the other hand, the expression for the purity at the bifurcation point conducts to an indeterminacy. Then, the analysis at this point needs to be conducted in a slightly different manner. To calculate the purity at the bifurcation point, we need the asymptotic expressions for $U_{{}^\psi}^a$ and $U_{{}^\chi}^a$ for $\omega \to 0$ and $\alpha \to 0$, respectively, which are given by
\begin{subequations}
\begin{align}
    U_{{}^\psi}^a&\sim B_a\omega_a  \,,\\
    U_{{}^\chi}^a&\sim B_a\alpha_a \,.
\end{align}
\end{subequations}
Using these results, the purity of a single particle of the system at the bifurcation point is given by
\begin{align}
\mu_1(a_1) =\mu_2(a_1) \sim \frac{2\sqrt{B_1B_2}}{B_1+B_2}\,,
\end{align}
which is different from zero, unlike the previous case (different frequencies). Notice that if $B_1=B_2$, $\mu_1(a_1)=1$, which also holds for arbitrary parameters $(X,Y,Z)$. This means that the system is not entangled.


\section{Conclusions}\label{sec:conclu}

In this work, we have introduced a time-dependent extension of the quantum geometric tensor, defined on a time-parameter manifold, allowing for a full covariant description of the geometry of the parameter space in time-evolving quantum systems. By considering small variations in both time and wave function parameters, we have demonstrated that this tensor adds new temporal components, providing a new tool for systems where the quantum state is non-time-separable and/or explicitly time-dependent. In particular, we have shown that the time-time component of the tQGT is proportional to the energy dispersion of the system, whereas if we consider stationary states the tQGT reduces to the standard QGT. We must mention that there are other proposals to incorporate the time in the QGT; they do this by considering time-dependent states in the original formula of the QGT \cite{Rattacas2020, Ilia2024}. However, they do not introduce the time-time and time-parameter components that we obtain in the tQGT.

The real part of the tQGT defines a time-dependent quantum metric tensor that generalizes the standard QGT of Provost and Valle, and its imaginary part provides a time-dependent (tBerry) curvature that generalizes the Berry curvature. The tQGT facilitates a better understanding of the dynamical geometric responses of quantum systems. We must emphasize that this tensor could distinguish between positive and negative average energies. Additionally, we obtained information by considering all the parameters in the wave function—which are not necessarily present in the corresponding quantum Hamiltonian—as explained in the examples. This is not possible with the standard QGT.

To illustrate a practical application of the tQGT, we have first applied our theoretical framework to a system corresponding to a harmonic/inverted oscillator, then to a time-dependent harmonic oscillator, and finally to a chain of generalized harmonic/inverted oscillators. For the harmonic/inverted oscillator system, in both cases (harmonic and inverted oscillator), we have obtained a wave function dependent on the parameter $B$ (among other parameters), using the corresponding system's kernel from an initial wave packet. The resulting wave functions belong to the family of  Gaussian states \eqref{eq:funcionondagen}, for which we have found that tQMT and tBerry curvature satisfy the general properties \eqref{eq:propiedadesgenerales}. In particular, for this family of wave functions, the determinant of the tQMT is zero if the dimension of the time-parameter manifold is greater than two; for any 2-dimensional time-parameter manifolds the tQMT satisfies the Palumbo and Goldman relation and its scalar curvature is $-16$, meaning that the 2-dimensional time-parameter space of these Gaussian states possesses a hyperbolic geometry. 

For the harmonic oscillator regime, the analysis showed that when $B = 1$, the usual ground state of the harmonic oscillator is recovered. The expected value of the Hamiltonian was calculated, revealing that the parameter $B$ is associated with the system's energy. The minimum energy corresponds to $B = 1$, while for other values of $B$, the state is a combination of all even energy eigenstates. Then, we computed the tQGT, showing its dependence on $B$ and identifying the ground state by a minimum in specific tensor components. Additionally, we have found that the tBerry curvature is different from zero and oscillates in time, meaning that the corresponding probability distribution $\rho_H(q, t, B)$ oscillates between the two values of $B$ with the same energy.  On the other hand, we have computed the expected value of the Hamiltonian for the inverted oscillator and found that, for $B>0$, the energy changes from negative to positive at $B=1$. Also, at $B=1$, all the components of the tQMT became bounded. We have observed that for large $\omega t$, some components of the tQMT have a minimum at $B=1$ or in the neighborhood of this value. We have also analyzed the transition region between the harmonic and inverted oscillators, focusing on the changes in the system's properties. We have found that the tQMT and tBerry curvature components obtained by varying the parameter $X$ diverge at $X = 0$, reflecting a big change in the system. 

We have studied the time-dependent harmonic oscillator to illustrate our formalism for time-dependent systems. We have shown that the Palumbo-Goldman relation is modified for this system, as shown in \eqref{eq:Palumbdomodi}. We have found that, as $g_{00}$ is time-independent, the energy dispersion of the system remains constant. Additionally, the system suffers a transition from an oscillator to a free particle in the limit where the time-frequency approaches zero. In our particular case, two choices for the parameters satisfy this condition. In one case, many metric components approach zero while others diverge. In contrast, in the other case, there is a metric whose determinant becomes constant, i.e., parameter-independent. Similarly, in this case, some components of the tBerry curvature become constant. This demonstrates that the tQGT effectively captures the relevant physical information associated with parameter variation.

We have investigated the generalized wave functions \eqref{eq:funciondeondaacoplados}, focusing on the tQMT and tBerry curvature. We have demonstrated that the determinant of the metric vanishes under specific conditions, particularly when the number of independent parameters exceeds $2N - l$, where $l$ denotes the multiplicity of functional dependencies among the $U^a$ and $V^a$ functions. We then considered metrics derived from varying a subset of parameters. The scalar curvature for these metrics revealed interesting patterns, particularly for $N = 2$, where we have found $\mathcal{R}[\Lambda^{I_1},\dots,\Lambda^{I_{4}}]=-32$ across different parameter sets. This pattern suggested a broader property that may extend to higher dimensions (\(N > 2\)). Although we did not provide formal proof, the consistency observed in multiple specific examples supports the hypothesis that the scalar curvature satisfies $\mathcal{R}[\Lambda^{I_1},\dots,\Lambda^{I_{2N}}]=-16N$ for any set of functionally independent $U^a$ and $V^a$.

One system that admits solutions of the form \eqref{eq:funciondeondaacoplados} is the generalized chain described by \eqref{hamiltonianoqpyfija}. In the case $N=2$, we identified three distinct regions, which, in normal coordinates correspond to the following: in region 1, two harmonic oscillators; in region 2, one inverted and one harmonic oscillator; and region 3, two inverted oscillators. We computed the tQMT and tBerry curvature for the three regions. Our results revealed that the scalar curvature is highly sensitive to variations in parameters $B_1$ and $B_2$ for large values of the time parameter $t$. Notably, this sensitivity is pronounced near $(B_1, B_2)=(1,1)$, indicating that the system undergoes the most abrupt changes in curvature in this region. As we have shown, this point corresponds to the ground state of the system. 

On the other hand, even when the states we are working with are non-time-separable, they are Gaussian. This allowed us to derive the purity of the reduced system by analyzing the covariance matrices in the generalized chain case. We obtained the purity of the $k$ oscillators of $N$ in region $m$. We discussed in detail the case of one oscillator in a two-oscillator system across the three regions, considering different frequencies. Our study showed that at the two bifurcation points, the purity of the reduced system drops to zero, indicating complete entanglement. Furthermore, we have shown that, as the state is a combination of even excited energy-eigenstates, the purity of the reduced system of one oscillator embedded in a system of two oscillators is zero at the bifurcation points for any even excited energy-eigenstate.

To close, we think that all components of the tQGT admit direct experimental access. In particular, the purely parameter elements \(Q_{ij}\) could be obtained by standard QGT measurement protocols (mentioned in the introduction), while the mixed temporal–parameter entries \(Q_{0i}\) and the purely temporal element \(Q_{00}\) could be extracted by applying a sudden Hamiltonian quench and monitoring the ensuing transition probability or energy-variance. Furthermore, since the states considered here are Gaussian, we could directly compute other quantities related to quantum entanglement beyond purity. For example, logarithmic negativity, generalized purities, Bastiaans-Tsallis entropies, and Rényi entropies \cite{renyi1970} \cite{serafini2017, Bastiaans84, Tsallis88, adesso2014, Diaz2023}. Finally, our formalism has several potential applications, such as out-of-equilibrium, dissipative, or non-Hermitian systems. In the case of dissipative or out-of-equilibrium systems \cite{Rattacas2020, Verma}, our approach could provide new information, since the previous work does not consider the new components of the tQGT that we introduce in our article. The case of non-Hermitian systems requires special attention, since the modification introduced in the inner product \cite{Bender}, with the time-inversion operator, most likely introduces additional terms in the tQGT, which must be taken into account to have an appropriate definition for these cases.

\section*{Acknowledgments}
This work was partially supported by DGAPA-PAPIIT Grants \break No. IN105422 and IN114225, also by the Spanish Ministerio de Ciencia Innovación y Universidades - Agencia Estatal de Investigación grant AEI/PID2020116567GB-C22. B. D. acknowledges support from SECIHTI (M\'exico)  No 371778. M. J. H. acknowledges support from CONAHCyT (M\'exico)  No 940148. Diego Gonzalez acknowledges the financial support of Instituto Polit\'ecnico Nacional, Grant No. SIP-20253696, and the postdoctoral fellowship from Consejo Nacional de Humanidades, Ciencia y Tecnología (CONAHCyT), México.
\appendix

\section*{References}
\bibliographystyle{iopart-num}
\bibliography{BIBTeX_File}

@article{Palumbo,
  title = {Revealing Tensor Monopoles through Quantum-Metric Measurements},
  author = {Palumbo, Giandomenico and Goldman, Nathan},
  journal = {Phys. Rev. Lett.},
  volume = {121},
  issue = {17},
  pages = {170401},
  numpages = {6},
  year = {2018},
  month = {Oct},
  publisher = {American Physical Society},
  doi = {10.1103/PhysRevLett.121.170401},
  url = {https://link.aps.org/doi/10.1103/PhysRevLett.121.170401}
}

@article{AA3,
  title = {Geometry of quantum evolution},
  author = {Anandan, J. and Aharonov, Y.},
  journal = {Phys. Rev. Lett.},
  volume = {65},
  issue = {14},
  pages = {1697--1700},
  numpages = {0},
  year = {1990},
  month = {Oct},
  publisher = {American Physical Society},
  doi = {10.1103/PhysRevLett.65.1697},
  url = {https://link.aps.org/doi/10.1103/PhysRevLett.65.1697}
}

@article{Juarez,
doi = {10.1088/1402-4896/aceb21},
url = {https://dx.doi.org/10.1088/1402-4896/aceb21},
year = {2023},
month = {aug},
publisher = {IOP Publishing},
volume = {98},
number = {9},
pages = {095106},
author = {Sergio B Juárez and Diego Gonzalez and Daniel Gutiérrez-Ruiz and J David Vergara},
title = {Generalized quantum geometric tensor for excited states using the path integral approach},
journal = {Physica Scripta}
}

@article{AA1,
  title = {Phase change during a cyclic quantum evolution},
  author = {Aharonov, Y. and Anandan, J.},
  journal = {Phys. Rev. Lett.},
  volume = {58},
  issue = {16},
  pages = {1593--1596},
  numpages = {0},
  year = {1987},
  month = {Apr},
  publisher = {American Physical Society},
  doi = {10.1103/PhysRevLett.58.1593},
  url = {https://link.aps.org/doi/10.1103/PhysRevLett.58.1593}
}

@article{AA2,
  title = {Geometric quantum phase and angles},
  author = {Anandan, J. and Aharonov, Y.},
  journal = {Phys. Rev. D},
  volume = {38},
  issue = {6},
  pages = {1863--1870},
  numpages = {0},
  year = {1988},
  month = {Sep},
  publisher = {American Physical Society},
  doi = {10.1103/PhysRevD.38.1863},
  url = {https://link.aps.org/doi/10.1103/PhysRevD.38.1863}
}

@article{Samuel,
  title = {General Setting for Berry's Phase},
  author = {Samuel, Joseph and Bhandari, Rajendra},
  journal = {Phys. Rev. Lett.},
  volume = {60},
  issue = {23},
  pages = {2339--2342},
  numpages = {0},
  year = {1988},
  month = {Jun},
  publisher = {American Physical Society},
  doi = {10.1103/PhysRevLett.60.2339},
  url = {https://link.aps.org/doi/10.1103/PhysRevLett.60.2339}
}

@article{Samuel1,
  title = {Observation of topological phase by use of a laser interferometer},
  author = {Bhandari, Rajendra and Samuel, Joseph},
  journal = {Phys. Rev. Lett.},
  volume = {60},
  issue = {13},
  pages = {1211--1213},
  numpages = {0},
  year = {1988},
  month = {Mar},
  publisher = {American Physical Society},
  doi = {10.1103/PhysRevLett.60.1211},
  url = {https://link.aps.org/doi/10.1103/PhysRevLett.60.1211}
}

@article{Pancha,
	author = {Pancharatnam, S. },
	date = {1956/11/01},
	doi = {10.1007/BF03046050},
	id = {Pancharatnam1956},
	isbn = {0370-0089},
	journal = {Proceedings of the Indian Academy of Sciences - Section A},
	number = {5},
	pages = {247--262},
	title = {Generalized theory of interference, and its applications},
	url = {https://doi.org/10.1007/BF03046050},
	volume = {44},
	year = {1956}}

@article{Osinv01,
title = {The Riemann Zeta Function and the Inverted Harmonic Oscillator},
journal = {Annals of Physics},
volume = {254},
number = {1},
pages = {25-40},
year = {1997},
issn = {0003-4916},
doi = {10.1006/aphy.1996.5636},
url = {https://www.sciencedirect.com/science/article/pii/S0003491696956365},
author = {R.K. Bhaduri and Avinash Khare and S.M. Reimann and E.L. Tomusiak}
}

@article{Osinv1,
  title = {Chaos and complexity in quantum mechanics},
  author = {Ali, Tibra and Bhattacharyya, Arpan and Haque, S. Shajidul and Kim, Eugene H. and Moynihan, Nathan and Murugan, Jeff},
  journal = {Phys. Rev. D},
  volume = {101},
  issue = {2},
  pages = {026021},
  numpages = {11},
  year = {2020},
  month = {Jan},
  publisher = {American Physical Society},
  doi = {10.1103/PhysRevD.101.026021},
  url = {https://link.aps.org/doi/10.1103/PhysRevD.101.026021}
}

@article{Osinv2,
	author = {Bhattacharyya, Arpan and Haque, S. Shajidul and Kim, Eugene H.},
	date = {2021/10/04},
	doi = {10.1007/JHEP10(2021)028},
	id = {Bhattacharyya2021},
	isbn = {1029-8479},
	journal = {Journal of High Energy Physics},
	number = {10},
	pages = {28},
	title = {Complexity from the reduced density matrix: a new diagnostic for chaos},
	url = {https://doi.org/10.1007/JHEP10(2021)028},
	volume = {2021},
	year = {2021}}

@Article{Osinv3,
	title={{The multi-faceted inverted harmonic oscillator: Chaos and complexity}},
	author={Arpan Bhattacharyya and Wissam Chemissany and S. Shajidul Haque and Jeff Murugan and Bin Yan},
	journal={SciPost Phys. Core},
	volume={4},
	pages={002},
	year={2021},
	publisher={SciPost},
	doi={10.21468/SciPostPhysCore.4.1.002},
	url={https://scipost.org/10.21468/SciPostPhysCore.4.1.002},
}

@article{Nepo,
  title = {Magnetic monopoles from antisymmetric tensor gauge fields},
  author = {Nepomechie, Rafael I.},
  journal = {Phys. Rev. D},
  volume = {31},
  issue = {8},
  pages = {1921--1924},
  numpages = {0},
  year = {1985},
  month = {Apr},
  publisher = {American Physical Society},
  doi = {10.1103/PhysRevD.31.1921},
  url = {https://link.aps.org/doi/10.1103/PhysRevD.31.1921}
}

@article{Freund,
title = {Dynamics of dimensional reduction},
journal = {Physics Letters B},
volume = {97},
number = {2},
pages = {233-235},
year = {1980},
issn = {0370-2693},
doi = {10.1016/0370-2693(80)90590-0},
url = {https://www.sciencedirect.com/science/article/pii/0370269380905900},
author = {Peter G.O. Freund and Mark A. Rubin}
}

@article{Tong,
  title = {Kinematic Approach to the Mixed State Geometric Phase in Nonunitary Evolution},
  author = {Tong, D. M. and Sj\"oqvist, E. and Kwek, L. C. and Oh, C. H.},
  journal = {Phys. Rev. Lett.},
  volume = {93},
  issue = {8},
  pages = {080405},
  numpages = {4},
  year = {2004},
  month = {Aug},
  publisher = {American Physical Society},
  doi = {10.1103/PhysRevLett.93.080405},
  url = {https://link.aps.org/doi/10.1103/PhysRevLett.93.080405}
}

@article{Marmo,
	author = {Chaturvedi, S. and Ercolessi, E. and Marmo, G. and Morandi, G. and Mukunda, N. and Simon, R.},
	date = {2004/06/01},
	doi = {10.1140/epjc/s2004-01814-5},
	id = {Chaturvedi2004},
	isbn = {1434-6052},
	journal = {The European Physical Journal C - Particles and Fields},
	number = {3},
	pages = {413--423},
	title = {Geometric phase for mixed states: a differential geometric approach},
	url = {https://doi.org/10.1140/epjc/s2004-01814-5},
	volume = {35},
	year = {2004}}

@article{Leyvraz,
  title = {Large-$N$ Scaling Behavior of the Lipkin-Meshkov-Glick Model},
  author = {Leyvraz, F. and Heiss, W. D.},
  journal = {Phys. Rev. Lett.},
  volume = {95},
  issue = {5},
  pages = {050402},
  numpages = {4},
  year = {2005},
  month = {Jul},
  publisher = {American Physical Society},
  doi = {10.1103/PhysRevLett.95.050402},
  url = {https://link.aps.org/doi/10.1103/PhysRevLett.95.050402}
}

@article{Caprio,
title = {Excited state quantum phase transitions in many-body systems},
journal = {Annals of Physics},
volume = {323},
number = {5},
pages = {1106-1135},
year = {2008},
issn = {0003-4916},
doi = {10.1016/j.aop.2007.06.011},
url = {https://www.sciencedirect.com/science/article/pii/S0003491607001042},
author = {M.A. Caprio and P. Cejnar and F. Iachello}
}

@article{Berry1984,
	ISSN = {00804630},
	URL = {http://www.jstor.org/stable/2397741},
	author = {M. V. Berry},
	journal = {Proc. R. Soc. A},
	number = {1802},
	pages = {45--57},
	publisher = {The Royal Society},
	title = {Quantal Phase Factors Accompanying Adiabatic Changes},
	volume = {392},
	year = {1984}
}

@article{Yuce,
	doi = {10.1088/0031-8949/74/1/014},
	url = {https://dx.doi.org/10.1088/0031-8949/74/1/014},
	year = {2006},
	month = {jun},
	publisher = {},
	volume = {74},
	number = {1},
	pages = {114},
	author = {C Yuce and A Kilic and A Coruh},
	title = {Inverted oscillator},
	journal = {Physica Scripta}
}

@article{Ali,
	title = {Chaos and complexity in quantum mechanics},
	author = {Ali, Tibra and Bhattacharyya, Arpan and Haque, S. Shajidul and Kim, Eugene H. and Moynihan, Nathan and Murugan, Jeff},
	journal = {Phys. Rev. D},
	volume = {101},
	issue = {2},
	pages = {026021},
	numpages = {11},
	year = {2020},
	month = {Jan},
	publisher = {American Physical Society},
	doi = {10.1103/PhysRevD.101.026021},
	url = {https://link.aps.org/doi/10.1103/PhysRevD.101.026021}
}

@book{Griffiths,
	title={Introduction to Quantum Mechanics},
	author={Griffiths, D.J.},
	isbn={9781107179868},
	lccn={2016427941},
	url={https://books.google.com.mx/books?id=0h-nDAAAQBAJ},
	year={2017},
	publisher={Cambridge University Press}
}

@article{Barton,
	title = {Quantum mechanics of the inverted oscillator potential},
	journal = {Annals of Physics},
	volume = {166},
	number = {2},
	pages = {322-363},
	year = {1986},
	issn = {0003-4916},
	doi = {10.1016/0003-4916(86)90142-9},
	url = {https://www.sciencedirect.com/science/article/pii/0003491686901429},
	author = {G. Barton}
}

@article{Osinv02,
title = {Physics of the Inverted Harmonic Oscillator: From the lowest Landau level to event horizons},
journal = {Annals of Physics},
volume = {435},
pages = {168470},
year = {2021},
note = {Special issue on Philip W. Anderson},
issn = {0003-4916},
doi = {10.1016/j.aop.2021.168470},
url = {https://www.sciencedirect.com/science/article/pii/S0003491621000762},
author = {Varsha Subramanyan and Suraj S. Hegde and Smitha Vishveshwara and Barry Bradlyn}
}

@article{Provost,
author="Provost, J. P.
and Vallee, G.",
title="Riemannian structure on manifolds of quantum states",
journal="Commun. Math. Phys.",
year="1980",
month="Sep",
day="01",
volume="76",
number="3",
pages="289--301",
issn="1432-0916",
doi="10.1007/BF02193559",
url="https://doi.org/10.1007/BF02193559"
}

@article{Carollo2020,
title = "Geometry of quantum phase transitions",
journal = "Phys. Rep.",
volume = "838",
pages = "1 - 72",
year = "2020",
issn = "0370-1573",
url = "http://www.sciencedirect.com/science/article/pii/S0370157319303655",
author = "Angelo Carollo and Davide Valenti and Bernardo Spagnolo"
}

@article{Zanardi2007Information,
  title = {Information-Theoretic Differential Geometry of Quantum Phase Transitions},
  author = {Zanardi, Paolo and Giorda, Paolo and Cozzini, Marco},
  journal = {Phys. Rev. Lett.},
  volume = {99},
  issue = {10},
  pages = {100603},
  numpages = {4},
  year = {2007},
  month = {Sep},
  publisher = {American Physical Society},
  doi = {10.1103/PhysRevLett.99.100603},
  url = {https://link.aps.org/doi/10.1103/PhysRevLett.99.100603}
}

@article{Alvarez2017,
	author="Alvarez-Jimenez, Javier
	and Dector, Aldo
	and Vergara, J. David",
	title="Quantum information metric and Berry curvature from a Lagrangian approach",
	journal="J. High Energy Phys.",
	year="2017",
	month="Mar",
	day="08",
	volume="2017",
	number="03",
	pages="044",
	issn="1029-8479",
	doi="10.1007/JHEP03(2017)044",
	url="https://doi.org/10.1007/JHEP03(2017)044"
}

@article{Sarkar2012,
  title = {Information geometry and quantum phase transitions in the Dicke model},
  author = {Dey, Anshuman and Mahapatra, Subhash and Roy, Pratim and Sarkar, Tapobrata},
  journal = {Phys. Rev. E},
  volume = {86},
  issue = {3},
  pages = {031137},
  numpages = {6},
  year = {2012},
  month = {Sep},
  publisher = {American Physical Society},
  doi = {10.1103/PhysRevE.86.031137},
  url = {https://link.aps.org/doi/10.1103/PhysRevE.86.031137}
}

@article{Gonzalez2020,
	url = {https://doi.org/10.1088/1751-8121/abc6c2},
	year = 2020,
	month = {nov},
	publisher = {{IOP} Publishing},
	volume = {53},
	number = {50},
	pages = {505305},
	author = {Diego Gonzalez and Daniel Guti{\'{e}}rrez-Ruiz and J David Vergara},
	title = {Phase space formulation of the Abelian and non-Abelian quantum geometric tensor},
	journal = {J. Phys. A: Math. Theor.}
}

@book{chruscinski,
	title={Geometric Phases in Classical and Quantum Mechanics},
	author={Chruscinski, D. and Jamiolkowski, A.},
	isbn={9780817681760},
	series={},
	year={2012},
	volume = {36},
	publisher = {Springer Science \& Business Media, New York},
}

@article{Anandan_1990,
  title = {Geometry of quantum evolution},
  author = {Anandan, J. and Aharonov, Y.},
  journal = {Phys. Rev. Lett.},
  volume = {65},
  issue = {14},
  pages = {1697--1700},
  numpages = {0},
  year = {1990},
  month = {Oct},
  publisher = {American Physical Society},
  doi = {10.1103/PhysRevLett.65.1697},
  url = {https://link.aps.org/doi/10.1103/PhysRevLett.65.1697}
}

@article{Simon,
  title = {Holonomy, the Quantum Adiabatic Theorem, and Berry's Phase},
  author = {Simon, Barry},
  journal = {Phys. Rev. Lett.},
  volume = {51},
  issue = {24},
  pages = {2167--2170},
  numpages = {0},
  year = {1983},
  month = {Dec},
  publisher = {American Physical Society},
  doi = {10.1103/PhysRevLett.51.2167},
  url = {https://link.aps.org/doi/10.1103/PhysRevLett.51.2167}
}

@article{Dodonov_2002,
	doi = {10.1088/1464-4266/4/3/362},
	url = {https://doi.org/10.1088/1464-4266/4/3/362},
	year = 2002,
	month = {apr},
	volume = {4},
	number = {3},
	pages = {S98--S108},
	author = {V V Dodonov},
	title = {Purity- and entropy-bounded uncertainty relations for mixed quantum states},
	journal = {Journal of Optics B: Quantum and Semiclassical Optics},
}

@article{Demarie2018,
	doi = {10.1088/1361-6404/aaaad0},
	url = {https://doi.org/10.1088/1361-6404/aaaad0},
	year = 2018,
	month = {mar},
	publisher = {{IOP} Publishing},
	volume = {39},
	number = {3},
	pages = {035302},
	author = {Tommaso F Demarie},
	title = {Pedagogical introduction to the entropy of entanglement for {G}aussian states},
	journal = {European Journal of Physics}
}

@Inbook{deGosson2019,
author="de Gosson, Maurice",
title="On the Purity and Entropy of Mixed {G}aussian States",
bookTitle="Landscapes of Time-Frequency Analysis",
year="2019",
publisher="Springer International Publishing",
address="Cham",
pages="145--158",
isbn="978-3-030-05210-2",
doi="10.1007/978-3-030-05210-2_5",
url="https://doi.org/10.1007/978-3-030-05210-2_5"
}

@book{deGosson2006,
  title={Symplectic Geometry and Quantum Mechanics},
  author={de Gosson, M.A.},
  isbn={9783764375751},
  lccn={2016346359},
  series={Operator Theory: Advances and Applications},
  url={https://doi.org/10.1007/3-7643-7575-2},
  doi={10.1007/3-7643-7575-2},
  year={2006},
  publisher={Birkh{\"a}user Basel}
}

@book{serafini2017,
  title={Quantum Continuous Variables: A Primer of Theoretical Methods},
  author={Serafini, A.},
  isbn={9781482246346},
  lccn={2016058596},
  year={2017},
  publisher={CRC Press, Taylor \& Francis Group}
}

@article{Golubeva2014,
author = {Golubeva, T. and Golubev, Yu.},
title = {Purity and Covariance Matrix},
journal = {Journal of Russian Laser Research},
volume = {35},
pages = {47-55},
doi = {10.1007/s10946-014-9399-2},
url = {https://doi.org/10.1007/s10946-014-9399-2},
year = {2014}
}

@book{diosi2011,
  title={A Short Course in Quantum Information Theory: An Approach From Theoretical Physics},
  author={Diosi, L.},
  isbn={9783642161179},
  series={Lecture Notes in Physics},
  url={https://books.google.es/books?id=w13vCAAAQBAJ},
  year={2011},
  publisher={Springer Berlin Heidelberg}
}

@article{Paris2003,
  title = {Purity of {G}aussian states: Measurement schemes and time evolution in noisy channels},
  author = {Paris, Matteo G. A. and Illuminati, Fabrizio and Serafini, Alessio and De Siena, Silvio},
  journal = {Phys. Rev. A},
  volume = {68},
  issue = {1},
  pages = {012314},
  numpages = {9},
  year = {2003},
  month = {Jul},
  publisher = {American Physical Society},
  doi = {10.1103/PhysRevA.68.012314},
  url = {https://link.aps.org/doi/10.1103/PhysRevA.68.012314}
}

@article{Agarwal1971,
  title = {Entropy, the {W}igner Distribution Function, and the Approach to Equilibrium of a System of Coupled Harmonic Oscillators},
  author = {Agarwal, G. S.},
  journal = {Phys. Rev. A},
  volume = {3},
  issue = {2},
  pages = {828--831},
  numpages = {0},
  year = {1971},
  month = {Feb},
  publisher = {American Physical Society},
  doi = {10.1103/PhysRevA.3.828},
  url = {https://link.aps.org/doi/10.1103/PhysRevA.3.828}
}

@article{Holevo1999,
  title = {Capacity of quantum {G}aussian channels},
  author = {Holevo, A. S. and Sohma, M. and Hirota, O.},
  journal = {Phys. Rev. A},
  volume = {59},
  issue = {3},
  pages = {1820--1828},
  numpages = {0},
  year = {1999},
  month = {Mar},
  publisher = {American Physical Society},
  doi = {10.1103/PhysRevA.59.1820},
  url = {https://link.aps.org/doi/10.1103/PhysRevA.59.1820}
}

@book{renyi1970,
  title={Probability Theory},
  author={R{\'{e}}nyi, A},
  year={1970},
   publisher={North-Holland Amsterdam}
}

@article{adesso2014,
author = {Adesso, Gerardo and Ragy, Sammy and Lee, Antony R.},
title = {Continuous Variable Quantum Information: {G}aussian States and Beyond},
journal = {Open Systems \& Information Dynamics},
volume = {21},
number = {01n02},
pages = {1440001},
year = {2014},
doi = {10.1142/S1230161214400010},

URL = { 
        https://doi.org/10.1142/S1230161214400010
    
}
}

@article{Bastiaans84,
author = {Martin J. Bastiaans},
journal = {J. Opt. Soc. Am. A},
number = {7},
pages = {711--715},
publisher = {Optica Publishing Group},
title = {New class of uncertainty relations for partially coherent light},
volume = {1},
month = {Jul},
year = {1984},
url = {http://opg.optica.org/josaa/abstract.cfm?URI=josaa-1-7-711},
doi = {10.1364/JOSAA.1.000711}
}

@article{Tsallis88,
author = {Tsallis, Constantino},
journal = {Journal of Statistical Physics},
pages = {479--487},
title = {New class of uncertainty relations for partially coherent light},
volume = {52},
year = {1988},
url = {https://doi.org/10.1007/BF01016429},
doi = {10.1007/BF01016429}
}

@article{Diaz2022,
  title = {Classical analogs of the covariance matrix, purity, linear entropy, and von Neumann entropy},
  author = {D\'{\i}az, Bogar and Gonz\'alez, Diego and Guti\'errez-Ruiz, Daniel and Vergara, J. David},
  journal = {Phys. Rev. A},
  volume = {105},
  issue = {6},
  pages = {062412},
  numpages = {14},
  year = {2022},
  month = {Jun},
  publisher = {American Physical Society},
  doi = {10.1103/PhysRevA.105.062412},
  url = {https://link.aps.org/doi/10.1103/PhysRevA.105.062412}
}

@article{Diaz2023,
  title = {Classical analogs of generalized purities, entropies, and logarithmic negativity},
  author = {D\'{\i}az, Bogar and Gonzalez, Diego and Hern\'andez, Marcos J. and Vergara, J. David},
  journal = {Phys. Rev. A},
  volume = {108},
  issue = {1},
  pages = {012411},
  numpages = {15},
  year = {2023},
  month = {Jul},
  publisher = {American Physical Society},
  doi = {10.1103/PhysRevA.108.012411},
  url = {https://link.aps.org/doi/10.1103/PhysRevA.108.012411}
}

@article{Lewis,
  title = {Classical and Quantum Systems with Time-Dependent Harmonic-Oscillator-Type Hamiltonians},
  author = {Lewis, H. R.},
  journal = {Phys. Rev. Lett.},
  volume = {18},
  issue = {13},
  pages = {510--512},
  numpages = {0},
  year = {1967},
  month = {Mar},
  publisher = {American Physical Society},
  doi = {10.1103/PhysRevLett.18.510},
  url = {https://link.aps.org/doi/10.1103/PhysRevLett.18.510}
}

@book{Dittrich, 
	title={Classical and Quantum Dynamics: From Classical Paths to Path Integrals},
	author={Dittrich, W. and Reuter, M.},
	isbn={9783319582986},
	year={2001},
	edition = {3rd},
	publisher={Springer}
}

@article{Yi2023,
  title = {Extracting the quantum geometric tensor of an optical Raman lattice by Bloch-state tomography},
  author = {Yi, Chang-Rui and Yu, Jinlong and Yuan, Huan and Jiao, Rui-Heng and Yang, Yu-Meng and Jiang, Xiao and Zhang, Jin-Yi and Chen, Shuai and Pan, Jian-Wei},
  journal = {Phys. Rev. Res.},
  volume = {5},
  issue = {3},
  pages = {L032016},
  numpages = {6},
  year = {2023},
  month = {Aug},
  publisher = {American Physical Society},
  doi = {10.1103/PhysRevResearch.5.L032016},
  url = {https://link.aps.org/doi/10.1103/PhysRevResearch.5.L032016}
}

@article{Yu2020,
    author = {Yu, Min and Yang, Pengcheng and Gong, Musang and Cao, Qingyun and Lu, Qiuyu and Liu, Haibin and Zhang, Shaoliang and Plenio, Martin B and Jelezko, Fedor and Ozawa, Tomoki and Goldman, Nathan and Cai, Jianming},
    title = {Experimental measurement of the quantum geometric tensor using coupled qubits in diamond},
    journal = {National Science Review},
    volume = {7},
    number = {2},
    pages = {254-260},
    year = {2019},
    month = {11},
    issn = {2095-5138},
    doi = {10.1093/nsr/nwz193},
    url = {https://doi.org/10.1093/nsr/nwz193},
    eprint = {https://academic.oup.com/nsr/article-pdf/7/2/254/38881668/nwz193.pdf},
}

@article{Ozawa2018,
  title = {Extracting the quantum metric tensor through periodic driving},
  author = {Ozawa, Tomoki and Goldman, Nathan},
  journal = {Phys. Rev. B},
  volume = {97},
  issue = {20},
  pages = {201117},
  numpages = {6},
  year = {2018},
  month = {May},
  publisher = {American Physical Society},
  doi = {10.1103/PhysRevB.97.201117},
  url = {https://link.aps.org/doi/10.1103/PhysRevB.97.201117}
}

@article{Rattacas2020,
doi = {10.1088/2399-6528/ab9505},
url = {https://dx.doi.org/10.1088/2399-6528/ab9505},
year = {2020},
month = {may},
publisher = {IOP Publishing},
volume = {4},
number = {5},
pages = {055017},
author = {Rattacaso, Davide and Vitale, Patrizia and Hamma, Alioscia},
title = {Quantum geometric tensor away from equilibrium},
journal = {Journal of Physics Communications}
}

@article{Ilia2024,
doi = {10.1038/s41467-024-48808-x},
url = {https://doi.org/10.1038/s41467-024-48808-x},
year = {2024},
month = {may},
volume = {15},
number = {1},
pages = {4621},
author = {Komissarov, Ilia and Holder, Tobias and Queiroz, Raquel},
title = {The quantum geometric origin of capacitance in insulators},
journal = {Nature Communications}
}

@article{Verma,
      title={Instantaneous Response and Quantum Geometry of Insulators}, 
      author={Nishchhal Verma and Raquel Queiroz},
      year={2025},
      eprint={2403.07052},
      archivePrefix={arXiv},
      primaryClass={cond-mat.mes-hall},
      url={https://arxiv.org/abs/2403.07052}, 
}

@book{Bender,
  title={Pt Symmetry: In Quantum And Classical Physics},
  author={Bender, C.M. and Dorey, P.E. and Dunning, T.C. and Fring, A. and Hook, D.W. and Jones, H.F. and Kuzhel, S. and Levai, G. and Tateo, R.},
  isbn={9781786345974},
  url={https://books.google.es/books?id=5PF9DwAAQBAJ},
  year={2018},
  publisher={World Scientific Publishing Company}
}
\end{document}